\documentclass[aip,jcp,reprint,numerical,amsmath,amssymb,amsfonts]{revtex4-1}
\usepackage{bm,graphicx,xcolor,hyperref,rotating,lineno,mathtools,multirow,makecell,placeins,physics}
\usepackage[version=4]{mhchem}
\usepackage{hyperref}
\hypersetup{
    colorlinks=true,
    linkcolor=blue,
    filecolor=blue,
    urlcolor=blue,
    citecolor=blue
}

\usepackage[normalem]{ulem}

\newcommand{\br}{\mathbf{r}}
\newcommand{\bp}{\mathbf{p}}
\newcommand{\bq}{\mathbf{q}}

\makeatletter
\newcommand\footnoteref[1]{\protected@xdef\@thefnmark{\ref{#1}}\@footnotemark}
\makeatother

\newcommand{\UWin}{Department of Chemistry, University of Winnipeg, Winnipeg, Manitoba, R3B 2E9, Canada}
\newcommand{\UMan}{Department of Chemistry, University of Manitoba, Winnipeg, Manitoba, R3T 2N2, Canada}

\begin{document}
%\linenumbers
\preprint{MBDatoms v1}

%	HEAD DATA
\title{Measuring correlated electron motion in atoms with the momentum-balance density}
\author{Lucy G. Todd}
\affiliation{\UWin}
\author{Joshua W. Hollett}
\email[Corresponding author: ]{j.hollett@uwinnipeg.ca}
\affiliation{\UWin}
\affiliation{\UMan}
\date{\today}

% ABSTRACT
\begin{abstract}
Three new measures of relative electron motion are introduced: equimomentum, antimomentum, and momentum-balance.  The equimomentum is the probability that two electrons have the exact same momentum, whereas the antimomentum is the probability their momenta are the exact opposite.  Momentum-balance (MB) is the difference between the equimomentum and antimomentum, and therefore indicates if equal or opposite momentum is more probably in a system of electrons.  The equimomentum, antimomentum and MB densities are also introduced, which are the local contribution to each quantity.  The MB and MB density of the extrapolated-Full Configuration Interaction wave functions of atoms of the first three rows of the periodic table are analyzed, with a particular focus on contrasting the correlated motion of electrons with opposite and parallel spin. Coulomb correlation between opposite-spin electrons leads to a higher probability of equimomentum, whereas Fermi correlation between parallel-spin electrons leads to a higher probability of antimomentum.  The local contribution to MB, given an electron is present, is a minimum at the nucleus and generally increases as the distance from the nucleus increases.  There are also interesting similarities between the effects of Fermi correlation and Coulomb correlation (of opposite-spin electrons) on MB.  
\end{abstract}

%	TITLE
\maketitle

%--------------------------
\section{Introduction}
%--------------------------
\label{sec:intro}

Most of the difficulties encountered in the study of electronic structure, both experimentally and computationally, arise from the complexity of the many-electron wave function. Therefore, both stand to benefit substantially from the unravelling this complexity.  The unravelling can begin by understanding the relative motion of electrons.  Not only does this help demystify the mesmerizing many-electron wave function, but it also lends itself to the development of more suitable models of electronic structure and a more complete picture of quantum mechanics.

Much has been learned about the relative position and momentum of electrons, and most is due to the analysis of intracules.\cite{Coulson1961, Lester1966, Coleman1967, Banyard1972, Thakkar1977, Banyard1978, Sharma1984, Thakkar1984, Boyd1988, Wang1992, Dominguez1992, Aguado1992, Sarasola1992, Banyard1993, Cann1993, CannJCP1993, CannIJQCS1993, Cioslowski1996, Fradera1997, Sarsa1998, Sarsa1999, Fradera2000, Koga2001, Koga2002, Koga2003, Gill2003TCA, Toulouse2007, Dumont2007, Crittenden2007, Crittenden2007omega, Bernard2008, Piris2008, Pearson2009, Per2009, RR11, Loos2010, Hollett2011Ecbond, Hollett2012, Bernard2013, Mercero2016, Hollett2017, RodriguezMayorga2018, RodriguezMayorga2019, ViaNadal2019} Intracules are two-electron probability distribution functions.\cite{Gill2006}  Constructed from the $N$-electron wave function (approximate or exact), these reduced probability densities reveal information regarding the relative position,\cite{Coulson1961, Banyard1972, Sharma1984, Koga2002, Hollett2012} momentum,\cite{Banyard1972, Banyard1978, Koga2002, Hollett2012} or position and momentum of electrons.\cite{Gill2003, Bernard2013}  The analyses often involve the calculation of intracule holes,\cite{Coulson1961,Cioslowski1998} which are the difference between the exact and the Hartree-Fock intracule.  An intracule hole explicitly describes the effects of Coulomb correlation\cite{Tew2007} on that particular intracule, and subsequent inferences are made regarding the correlated motion of the electrons.  Such studies have led to a better understanding of the relative motion of electrons due to radial and angular correlation in atoms,\cite{Banyard1972,Hollett2011Ecbond,Hollett2012} the correlated motion of electrons during bond dissociation,\cite{Hollett2011Ecbond} and their correlated motion in the presence of a static electric field.\cite{Hollett2017} However, for most intracules the spatial information ({\it i.e.}~absolute location of the electrons) is lost, which impedes the completion of a picture of relative electron motion in atoms and molecules.

By using a more complex object, such as the two-electron density, correlated electron movement can be analyzed in real space through the construction of Coulomb,\cite{Sanders1992, Banyard1993} exchange, or exchange-correlation\cite{Wang2010} holes.  In this context, the hole is the two-electron density as a function of the position of the second electron when a test position is chosen for the first.\cite{Cooper1978,Tew2007}  Such an approach can be used to demonstrate Fermi correlation and the Fermi hole, which occur between electrons of parallel-spin and is due to the antisymmetry of the electronic wave function.\cite{Tew2007}  This approach is also used to design and test models for density functional approximations.\cite{Baerends2001}  By choosing a position in the molecule for the test electron, the complexity of the two-electron density is reduced but a complete picture of the relative motion of the electrons must be synthesized from the combination of densities produced from a collection of test electron positions.  

Alternatively, there are intracule-like probability densities that are not reduced to a single (1D or 3D) variable.  Such probability densities, like the intex distribution,\cite{Proud2010,Proud2012,Valderrama2001} retain absolute position information, that is often lost with simpler intracules, and can provide information regarding the relative positions of electrons at different locations in an atom or molecule.  Similar distributions also exist for momentum space.\cite{Reed1980}

In an effort to unravel some of the complexity of the correlated motion of electrons this article presents multiple new quantum mechanical properties.  The properties are the equi- and antimomentum and momentum-balance, along with the corresponding equi- and antimomentum density and momentum-balance density.  The momentum-balance is a measure of the correlated motion in an electronic system, and the momentum-balance density describes the local contribution of the electronic wave function to the correlated electron motion.  In Section \ref{sec:theory}, expressions for the equimomentum, antimomentum, and momentum balance are derived, along with the corresponding densities.  This is followed by a description of the basic algorithm for the calculation of these properties in atoms and molecules using Gaussian basis sets.  Section \ref{sec:results} presents an analysis and discussion of the momentum-balance and momentum-balance density in atoms of the first 17 many-electron elements of the periodic table.  Then finally, the results are summarized and future applications of these properties are discussed in Section \ref{sec:conc}. Atomic units are used throughout.

%--------------------------
\section{Theory}
%--------------------------
\label{sec:theory}

The connection between electron position and momentum is through the Fourier transform.  The $N$-electron momentum-space wave function, $\Phi$, in terms of the position-space wave function, $\Psi$, is given as 
\begin{multline}
\label{eq:momwf}
\Phi(\bp_1,\bp_2,\dots,\bp_N) = \frac{1}{(2\pi)^{3N}} \int e^{-i\sum_{k=1}^N \bp_k \cdot \br_k}\\ \times \Psi(\br_1,\br_2,\dots,\br_N) d\br_1 d\br_2 \dots d\br_N,
\end{multline}
%\begin{equation}
%\label{eq:momwf}
%\Phi(\bp_1,\bp_2,\dots,\bp_N) = \frac{1}{(2\pi)^{3N}} \int e^{-i\sum_{k=1}^N \bp_k \cdot \br_k} \Psi(\br_1,\br_2,\dots,\br_N) d\br_1 d\br_2 \dots d\br_N,
%\end{equation}
where $\bp_k$ and $\br_k$ are the electron momenta and positions, respectively.  The wave functions are spinless, that is, they have been previously integrated with respect to the electron spin-coordinates.  When addressing the relative motion of electrons, it is simpler to focus on electron pairs rather than all $N$-electrons at once.  Therefore, the two-electron reduced density matrix (2-RDM), $\Gamma(\br_1,\br_2,\br_1',\br_2')$, is formed via integration over the $N$-electron wave function,
\begin{multline}
\label{eq:pos2rdm}
\Gamma(\br_1,\br_2,\br_1',\br_2') = \int \Psi^*(\br_1',\br_2',\dots,\br_N)\\ \times \Psi(\br_1,\br_2,\dots,\br_N) d\br_3 \dots d\br_N,
\end{multline}
%\begin{equation}
%\label{eq:pos2rdm}
%\Gamma(\br_1,\br_2,\br_1',\br_2') = \int \Psi^*(\br_1',\br_2',\dots,\br_N) \Psi(\br_1,\br_2,\dots,\br_N) d\br_3 \dots d\br_N,
%\end{equation}
where the diagonal of the 2-RDM is the two-electron density, $\Gamma(\br_1,\br_2) = \Gamma(\br_1,\br_2,\br_1,\br_2)$.  Following Equation \eqref{eq:momwf}, the two-electron momentum density, $\Pi(\bp_1,\bp_2)$, can be formed from the position-space 2-RDM,
\begin{multline}
\label{eq:2emomden}
\Pi(\bp_1,\bp_2) = \frac{1}{(2\pi)^6} \int e^{i\left[\bp_1\cdot(\br_1'-\br_1) + \bp_2\cdot(\br_2'-\br_2) \right]} \\ \times \Gamma(\br_1,\br_2,\br_1',\br_2') d\br_1 d\br_2 d\br_1' d\br_2'.
\end{multline}
%\begin{equation}
%\label{eq:2emomden}
%\Pi(\bp_1,\bp_2) = \frac{1}{(2\pi)^6} \int e^{i\left[\bp_1\cdot(\br_1'-\br_1) + \bp_2\cdot(\br_2'-\br_2) \right]} \Gamma(\br_1,\br_2,\br_1',\br_2') d\br_1 d\br_2 d\br_1' d\br_2'.
%\end{equation}
The two-electron momentum density gives the simultaneous probability that one electron has momentum $\bp_1$ and another has $\bp_2$.  By inserting a Dirac delta function, $\delta(x)$, and using the identity, $\delta(x) = \frac{1}{2\pi}\int e^{-i k x} dk$, the probability that two electrons have the exact same momentum can be calculated,  
\begin{align}
\label{eq:equimom}
\lambda^+	= & \int \Pi(\bp_1,\bp_2) \delta(\bp_1 - \bp_2) d\bp_1 d\bp_2 \nonumber \\
		= & \frac{1}{(2\pi)^3} \int \Gamma(\br_1,\br_2,\br_1+\bq,\br_2 - \bq)  d\bq d\br_1 d\br_2,
\end{align}
which will be referred to as the {\it equimomentum}. Similarly, the probability that two electrons have the exact opposite momentum is given by,
\begin{align}
\label{eq:antimom}
\lambda^-	= & \int \Pi(\bp_1,\bp_2) \delta(\bp_1 + \bp_2) d\bp_1 d\bp_2 \nonumber \\
		= & \frac{1}{(2\pi)^3} \int \Gamma(\br_1,\br_2,\br_1+\bq,\br_2 + \bq)  d\bq d\br_1 d\br_2,
\end{align}
which will be referred to as the {\it antimomentum}. The difference between the two, will be referred to as the {\it momentum-balance (MB)},
\begin{equation}
\label{eq:mombal}
\mu = \lambda^+ - \lambda^-
\end{equation}
The MB is positive for a system in which it is more probable that the electrons have the same momentum, negative for a system in which opposite momenta are more probable, and zero when both scenarios are equally likely.

If the integrand for the equi- or antimomentum [Equation \eqref{eq:equimom} or \eqref{eq:antimom}] is integrated over only one of the electron coordinates the result is the local contribution to each quantity,
\begin{equation}
\label{eq:equiantiden}
\lambda^\pm(\br) =   \frac{1}{(2\pi)^3} \int \Gamma(\br,\br_2,\br+\bq,\br_2 \mp \bq) d\br_2 d\bq,
\end{equation}
the equimomentum density, $\lambda^+(\br)$, and the antimomentum density, $\lambda^-(\br)$.  The difference between the two is the MB density,
\begin{equation}
\label{eq:MBden}
\mu(\br) =  \lambda^+(\br) - \lambda^-(\br)
\end{equation}
Although $\lambda^+(\br)$ and $\lambda^-(\br)$ are not probability densities, the MB density can still be analyzed to reveal the regions of a wave function that contribute to electrons having equal or opposite momenta.  Such analysis is analogous to that seen with various types of energy densities.\cite{Ayers2002,Anderson2010,Bacskay2017}  The sign of the MB density reveals whether there is a local contribution to equi- or antimomentum,
\begin{align}
\mu(\br) > 0 \: & \implies & \text{contributes to {\bf equal} momenta} \nonumber \\
\mu(\br) < 0 \: & \implies & \text{contributes to {\bf opposite} momenta} \nonumber
\end{align}

Inspection of the definitions of $\lambda^+(\br)$ and $\lambda^-(\br)$ [Equation \eqref{eq:equiantiden}] in terms of the 2-RDM reveals a strong dependence of those densities and the MB density on the electron density at $\br$.  Therefore, a reduced MB density, $m(\br)$, is defined (similar to an energy potential) to eliminate that dependence,
\begin{equation}
	m(\br) = \frac{\mu(\br)}{\rho(\br)},
\end{equation}
where $\rho(\br)$ is the one-electron density. It is expected that $m(\br)$ will reveal more fine details of the relative momenta of electrons in a system in comparison to $\mu(\br)$.

%--------------------------
%\subsection{Recurrence Relations}
%--------------------------
%\label{subsec:RRs}

For wave functions constructed from an orbital basis, the equi- and antimomentum can be calculated by contracting the 2-RDM in the given basis, $\Gamma_{abcd}$, with the appropriate two-electron integrals,
\begin{equation}
\label{eq:equianticont}
\lambda^\pm = \sum_{abcd} \Gamma_{abcd} \left[abcd\right]_{\lambda^\pm}.
\end{equation}
The two-electron integrals are given by
\begin{multline}
\left[abcd\right]_{\lambda^\pm} =  \frac{1}{(2\pi)^3} \int \phi_a(\br_1) \phi_b(\br_1 + \bq) \\ \times \phi_c(\br_2) \phi_d(\br_2 \mp \bq) d\bq d\br_1 d\br_2,
\end{multline}
%\begin{equation}
%\left[abcd\right]_{\lambda^\pm} =  \frac{1}{(2\pi)^3} \int \phi_a(\br_1) \phi_b(\br_1 + \bq) \phi_c(\br_2) \phi_d(\br_2 \mp \bq) d\bq d\br_1 d\br_2,
%\end{equation}
where the orbitals, $\phi_a$, are assumed to be real. The integrals exhibit the very practical identity,
\begin{equation}
	\label{eq:ident}
	\left[abcd\right]_{\lambda^-} = \left[abdc\right]_{\lambda^+}.
\end{equation}
The equi- and antimomentum densities can be calculated in the same manner, except the four-orbital integral is replaced by the product of an orbital and a three-orbital integral, 
\begin{equation}
\lambda^\pm(\br) = \sum_{abcd} \Gamma_{abcd} \phi_a(\br)\left[bcd\right]_{\lambda^\pm}.
\end{equation}
The three-orbital integral is given by,
\begin{multline}
\left[bcd\right]_{\lambda^\pm} =  \frac{1}{(2\pi)^3} \int \phi_b(\br + \bq)\\ \times \phi_c(\br_2) \phi_d(\br_2 \mp \bq) d\bq d\br_2
\end{multline}
%\begin{equation}
%\left[bcd\right]_{\lambda^\pm} =  \frac{1}{(2\pi)^3} \int \phi_b(\br + \bq) \phi_c(\br_2) \phi_d(\br_2 \mp \bq) d\bq d\br_2
%\end{equation}
which also has a useful identity,
\begin{equation}
	\left[bcd\right]_{\lambda^-} = \left[bdc\right]_{\lambda^+}.
\end{equation}
Recurrence relations for the practical calculation of the molecular integrals, $\left[abcd\right]_{\lambda^\pm}$ and $\left[bcd\right]_{\lambda^\pm}$, over Gaussian-type basis functions of arbitrary angular momentum are provided in the supplementary material. 
%--------------------------
%\subsection{Hartree-Fock momentum-balance}
%--------------------------
%\label{subsec:HFmb}

Analysis of the relative motion of electrons via MB and the MB density, is aided by decomposing the 2-RDM into its various spin-components,
\begin{align}
	\Gamma(\br_1,\br_2,\br_1',\br_2') = & \Gamma^{\alpha\alpha}(\br_1,\br_2,\br_1',\br_2') + \Gamma^{\beta\beta}(\br_1,\br_2,\br_1',\br_2') \nonumber \\ & \Gamma^{\alpha\beta}(\br_1,\br_2,\br_1',\br_2') + \Gamma^{\beta\alpha}(\br_1,\br_2,\br_1',\br_2'),
\end{align}
where $\alpha$ denotes ``spin-up" and $\beta$ denotes ''spin-down".  To demonstrate the relationship between $\mu$ and correlated electron motion, consider the Hartree-Fock wave function.  Inserting the spin-resolved HF 2-RDM into Equation \eqref{eq:equianticont} yields an expression for the HF equi- and antimomentum,
\begin{multline}
	\lambda^\pm_\text{HF} = \frac{1}{2}\sum_{ij} n^\alpha_i n^\alpha_j \left( \left[i i j j \right]^{\alpha\alpha}_{\lambda^\pm} - \left[i j j i \right]^{\alpha\alpha}_{\lambda^\pm} \right) \\
+ n^\beta_i n^\beta_j \left( \left[i i j j \right]^{\beta\beta}_{\lambda^\pm} - \left[i j j i \right]^{\beta\beta}_{\lambda^\pm} \right) \\
+ n^\alpha_i n^\beta_j \left( \left[i i j j \right]^{\alpha\beta}_{\lambda^\pm} + \left[i i j j \right]^{\beta\alpha}_{\lambda^\pm} \right),
\end{multline}
where $n_i^\sigma$ is the occupancy of $\sigma$-spin orbital $i$, with $\sigma=\alpha$ or $\beta$. The integral identity relating equimomentum and antimomentum [Equation \eqref{eq:ident}] implies that the Coulomb-type integral for both quantities are equivalent,
\begin{equation}
	\left[i i j j \right]_{\lambda^-} = \left[i i j j \right]_{\lambda^+}.
\end{equation}
Considering this, and the decomposition of $\mu$ into spin-components,
\begin{equation}
	\mu_\text{HF} = \mu^{\alpha\alpha}_\text{HF} + \mu^{\beta\beta}_\text{HF} + \mu^{\alpha\beta}_\text{HF} + \mu^{\beta\alpha}_\text{HF},
\end{equation}
the following simplification can be made,
\begin{equation}
	\mu^{\sigma\sigma}_\text{HF} = \frac{1}{2}\sum_{ij} n^\sigma_i n^\sigma_j \left( \left[i j j i \right]^{\sigma\sigma}_{\lambda^-} - \left[i j j i \right]^{\sigma\sigma}_{\lambda^+} \right).
\end{equation}
The HF MB between electrons of parallel-spin is the difference between the equi- and antimomentum contributions that result from the exchange interaction ({\it i.e.} antisymmetry).  Given that there is no exchange interaction between electrons of opposite-spin, the HF MB of opposite-spin electrons is exactly zero,
\begin{equation}
	\mu^{\sigma\sigma'}_\text{HF} = 0.
\end{equation}
An analogous result is found for the MB density,
\begin{equation}
	\mu^{\sigma\sigma}_\text{HF}(\br) = \frac{1}{2}\sum_{ij} n^\sigma_i n^\sigma_j \phi_i^\sigma(\br) \left( \left[j j i \right]^{\sigma\sigma}_{\lambda^-} - \left[j j i \right]^{\sigma\sigma}_{\lambda^+} \right),
\end{equation}
and
\begin{equation}
	\mu^{\sigma\sigma'}_\text{HF}(\br) = 0.
\end{equation}

It is clear from this result that non-zero MB and MB density only exists when electron motion is correlated. In other words, if electron motion is uncorrelated then equimomentum and antimomentum are equally probable, $\lambda^+ = \lambda^-$ and $\lambda^+(\br) = \lambda^-(\br)$.  In the case of the HF wave function, correlation occurs only between parallel-spin electrons and is due to the antisymmetry of the wave function, often referred to as Fermi correlation.\cite{Tew2007}  Therefore, the MB and MB density of correlated ({\it e.g.} post-HF) wave functions reveal the presence of correlation between opposite-spin electrons and the change in correlation between parallel-spin electrons.

%-----------------------------------
\section{Method}
%-----------------------------------
\label{sec:method}

All wave functions (including 1 and 2-RDMs) were obtained using a determinant-driven selected configuration interaction (sCI) method known as CIPSI (Configuration Interaction using a Perturbative Selection made Iteratively)\cite{Huron1973,Giner2013,Giner2015} in which the energies are extrapolated to the full configuration interaction (FCI) result using multireference perturbation theory.\cite{Garniron2017,Loos2018,qp2}  The all-electron extrapolated-FCI (exFCI) calculations were performed using Quantum Package 2.0.\cite{qp2} The wave functions and RDMs for \ce{He} were obtained using the cc-pVTZ, aug-cc-pVTZ and aug-cc-pVQZ/f basis sets,\cite{Dunning1989,Woon1994} and all calculations beyond \ce{He} were performed using the aug-cc-pCVTZ/f basis set.\cite{Prascher2011,Feller1996,Schuchardt2007,BSE2019}

Using the RDMs provided by Quantum Package 2.0, the equimomentum, antimomentum, and MB, along with their densities were calculated using MUNgauss.\cite{MUNgauss}  The densities were calculated on a two-dimensional mesh, with a grid-spacing of 0.05 bohr, for visualization.  The densities were also numerically integrated on the SG-1 grid\cite{Gill1993} for comparison to analytically calculated equimomentum, antimomentum and MB values.

%-----------------------------------
\section{Results}
%-----------------------------------
\label{sec:results}

%-----------------------------------
\subsection{Opposite-spin correlation}
%-----------------------------------
\label{subsec:oppspin}

\begin{figure}
	\begin{center}
	\includegraphics[width=0.5\textwidth]{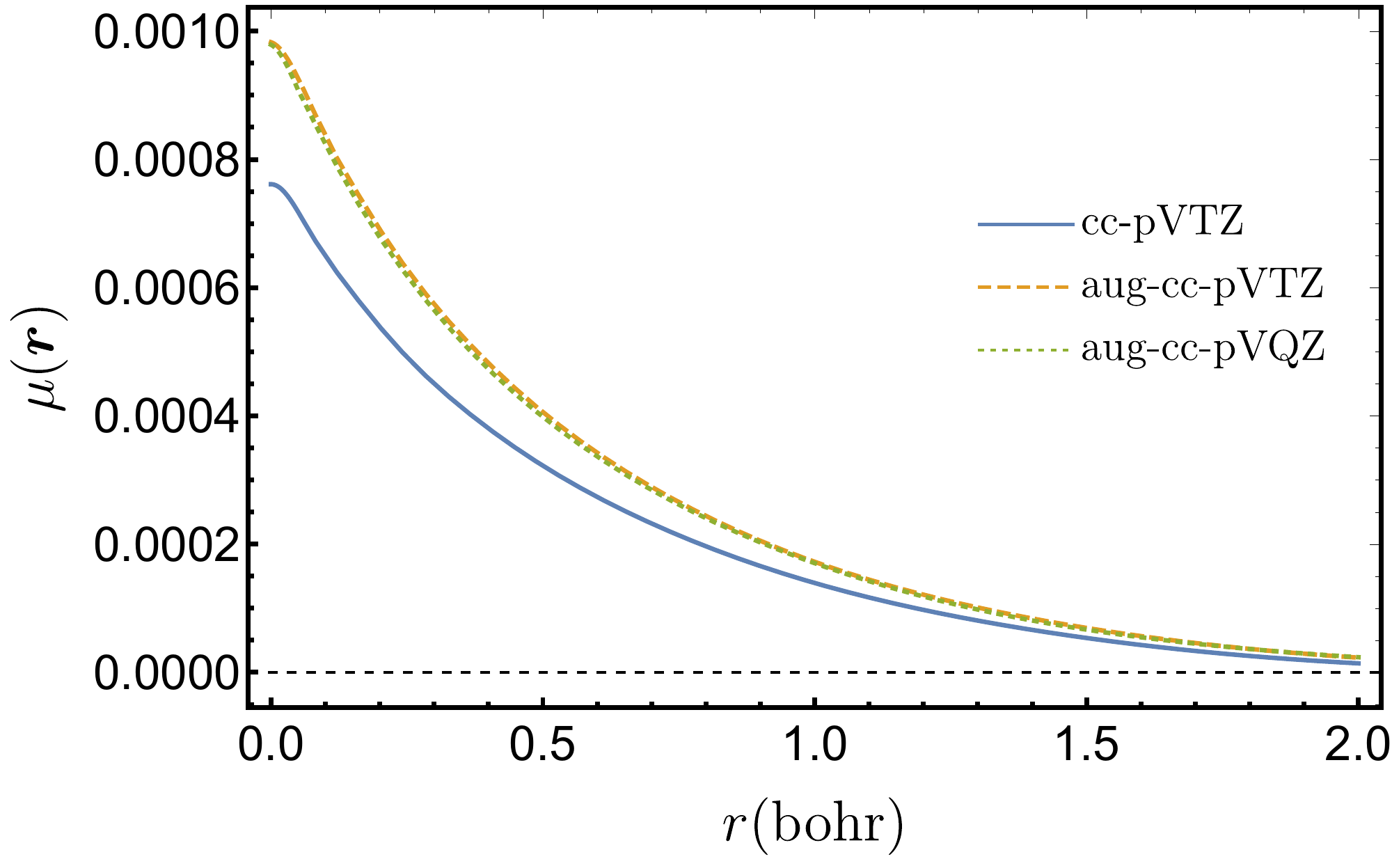}
	\caption{\label{fig:HeMBDbasis} Comparison of momentum-balance density of \ce{He} atom for multiple Dunning correlation-consistent basis sets.}
	\end{center}
\end{figure}

\begin{figure}
	\begin{center}
	\includegraphics[width=0.5\textwidth]{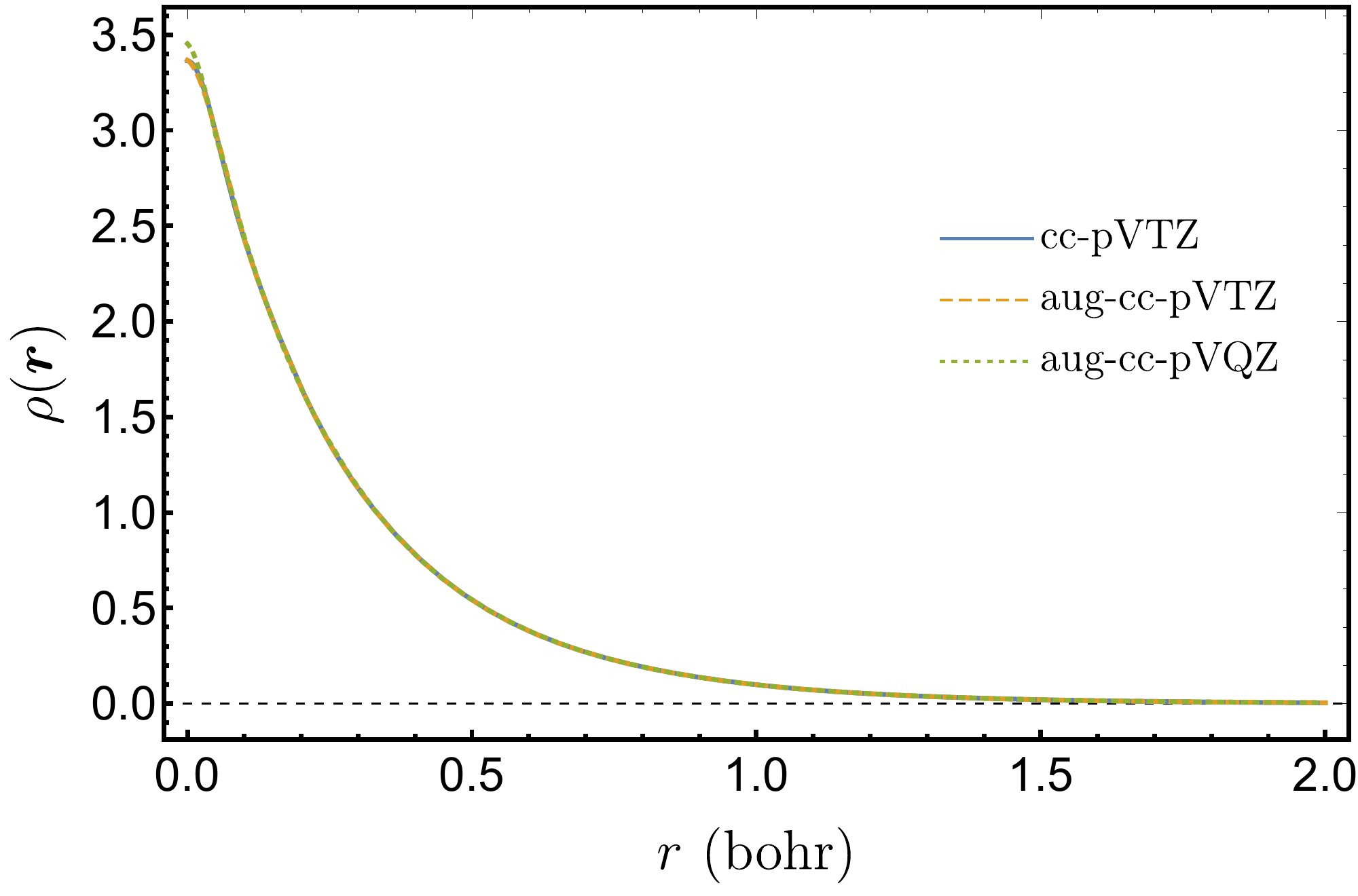}
	\caption{\label{fig:Herhobasis} Comparison of total one-electron density of \ce{He} atom for multiple Dunning correlation-consistent basis sets.}
	\end{center}
\end{figure}

As an initial assessment of basis set dependence, the MB density of the \ce{He} atom obtained using three different Dunning basis sets is shown in Figure \ref{fig:HeMBDbasis}.
It is seen that there is significant deviation between the MB density calculated using the cc-pVTZ basis set and the density calculated using the aug-cc-pVTZ basis set.  The deviation is most pronounced at the nucleus and decays as the distance from the nucleus increases. Therefore, it appears the addition of diffuse functions, or ``augmenting" the basis set, (cc-pVTZ to aug-cc-pVTZ) is more important to an accurate description of the MB density than adding an extra set of valence basis functions (aug-cc-pVTZ to aug-cc-pVQZ). Furthermore, the MB density is far more sensitive to basis set than the one-electron density (Figure \ref{fig:Herhobasis}).  This may not be surprising considering it is a comparison between a one-electron property and a two-electron property.  All subsequent properties of \ce{He} that follow are obtained using the aug-cc-pVQZ basis set.

\begin{figure}
	\begin{center}
	\includegraphics[width=0.5\textwidth]{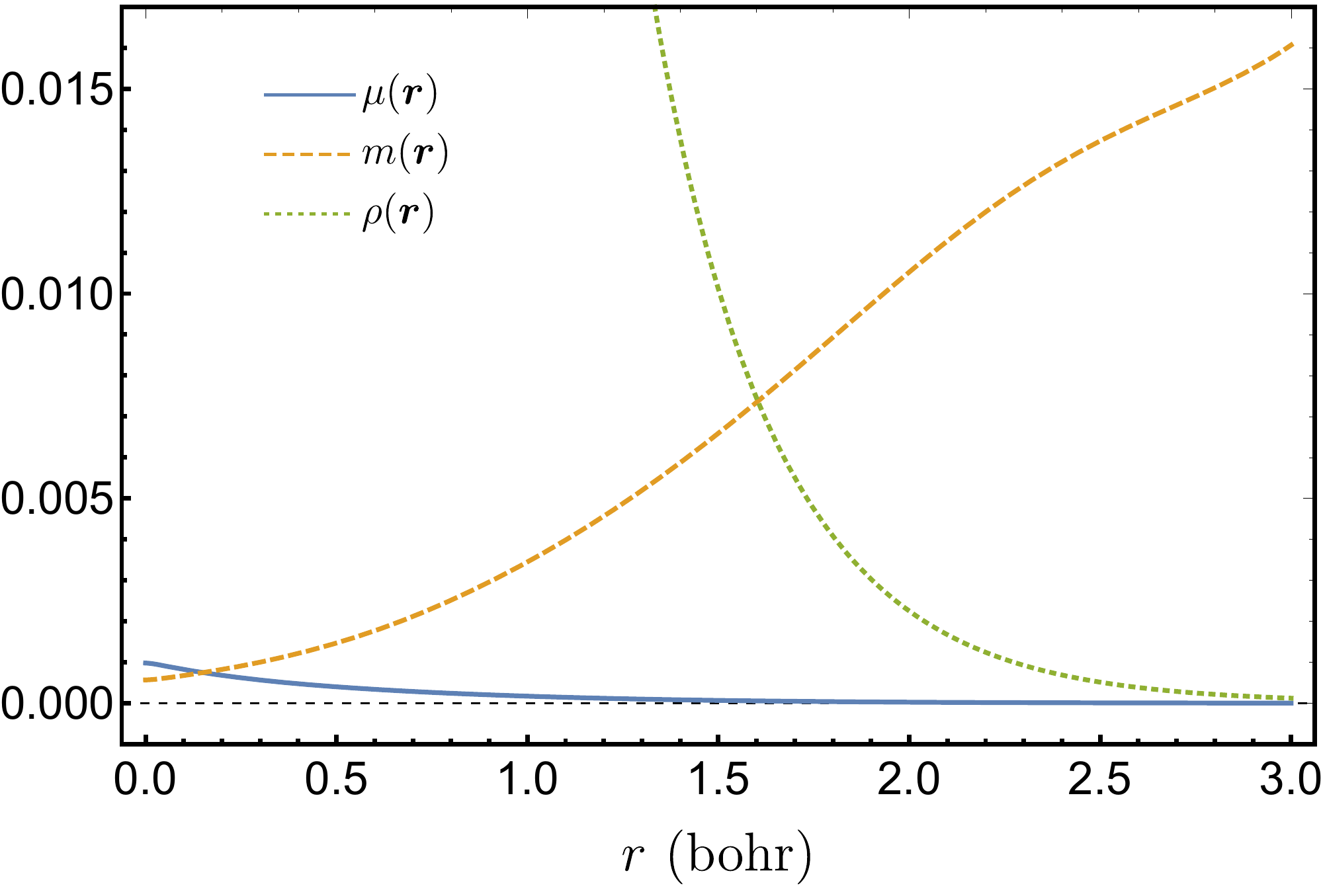}
	\caption{\label{fig:HeMBD} Total momentum-balance density and reduced momentum-balance density of \ce{He}, with the one-electron density.}
	\end{center}
\end{figure}

In Figure \ref{fig:HeMBD}, the \ce{He} atom MB density, $\mu(\br)$, reduced MB density, $m(\br)$, and one-electron density, $\rho(\br)$ are plotted together.
Evident from Figures \ref{fig:HeMBDbasis} and \ref{fig:HeMBD}, $\mu(\br)$ for \ce{He} is positive everywhere, and results in a total MB of 0.00373.  The MB is the difference between an equimomentum $\lambda^+ = 0.02481$ and an antimomentum $\lambda^- = 0.02107$. In other words, if one could measure the simultaneous momenta of the two electrons of ground state \ce{He} 100 000 times, they would find the electrons have the exact same momentum 2 481 times and the exact opposite momentum 2 107 times.  The difference, which is the MB, is due to Coulomb correlation between the $\alpha$ and $\beta$ electron.  Their correlated motion favours equimomentum over antimomentum, which is consistent with earlier studies of angular correlation in the \ce{He} atom.\cite{Banyard1978,Koga2003,Hollett2012}

From Figures \ref{fig:HeMBDbasis} and \ref{fig:Herhobasis} it is clear that $\mu(\br)$ closely follows the shape of the $\rho(\br)$, albeit with a much smaller magnitude and slower decay.  As mentioned previously, this is due to the dependence of the two-electron density on the one-electron density ({\it i.e.}~an electron must be present to interact).  This dependency is removed in the reduced MB density, $m(\br)$, which can be viewed as a MB potential. It is seen that $m(\br)$ is actually a minimum at the nucleus and the potential contribution to MB increases as the distance from the nucleus increases.  Of course, while $m(\br)$ continues to grow as $r$ increases, $\rho(\br)$ decays and eventually extinguishes the contribution.

\begin{figure}
	\begin{center}
	\includegraphics[width=0.5\textwidth]{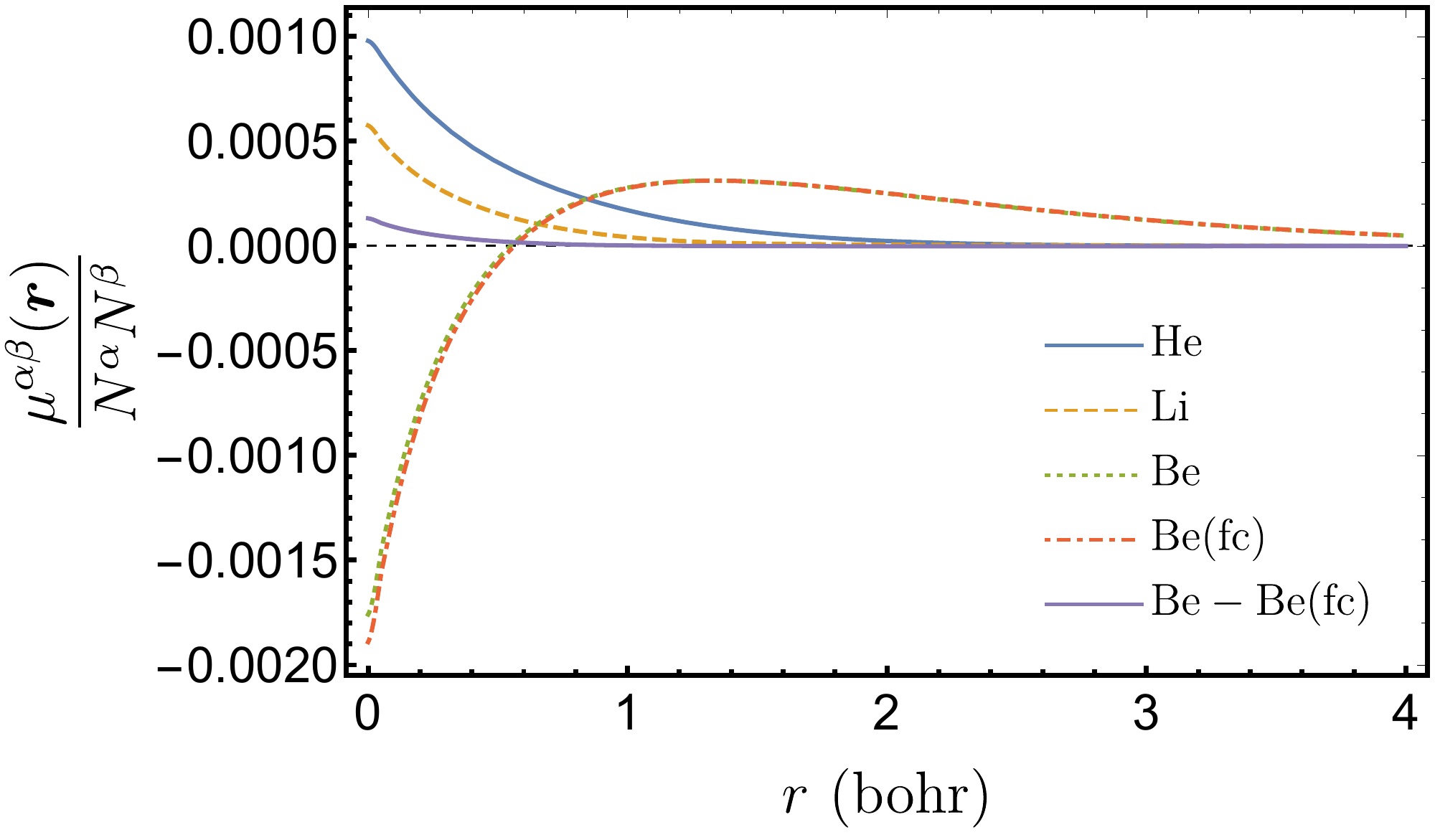}
	\caption{\label{fig:HetoBeMBD} Renormalized $\mu^{\alpha\beta}(\br)$ of \ce{He}, \ce{Li}, \ce{Be} and \ce{Be}(fc) [frozen core]. The difference between the all-electron and frozen core $\mu^{\alpha\beta}(\br)$ for \ce{Be} is included.}
	\end{center}
\end{figure}

Figure \ref{fig:HetoBeMBD} presents the $\alpha\beta$ MB density of \ce{He}, \ce{Li} and \ce{Be} divided by the number of $\alpha\beta$ electron pairs.
It also includes $\mu^{\alpha\beta}(\br)$ of \ce{Be} calculated using a frozen-core approximation to ascertain the role of the valence electrons in $\mu^{\alpha\beta}(\br)$. The frozen-core $\mu^{\alpha\beta}(\br)$ closely resembles the fully correlated $\mu^{\alpha\beta}(\br)$ which, unlike \ce{He} and \ce{Li}, has a node at small $r$ and a negative peak at the nucleus.  It is evident that the valence electrons are mainly responsible for the structure of $\mu^{\alpha\beta}(\br)$ for \ce{Be}.  The difference between the frozen-core and fully correlated $\mu^{\alpha\beta}(\br)$ reveals the contribution from the core, and core-valence, correlation and it is very similar to the $\mu^{\alpha\beta}(\br)$ of \ce{He} and \ce{Li}.

\begin{figure}
	\begin{center}
	\includegraphics[width=0.5\textwidth]{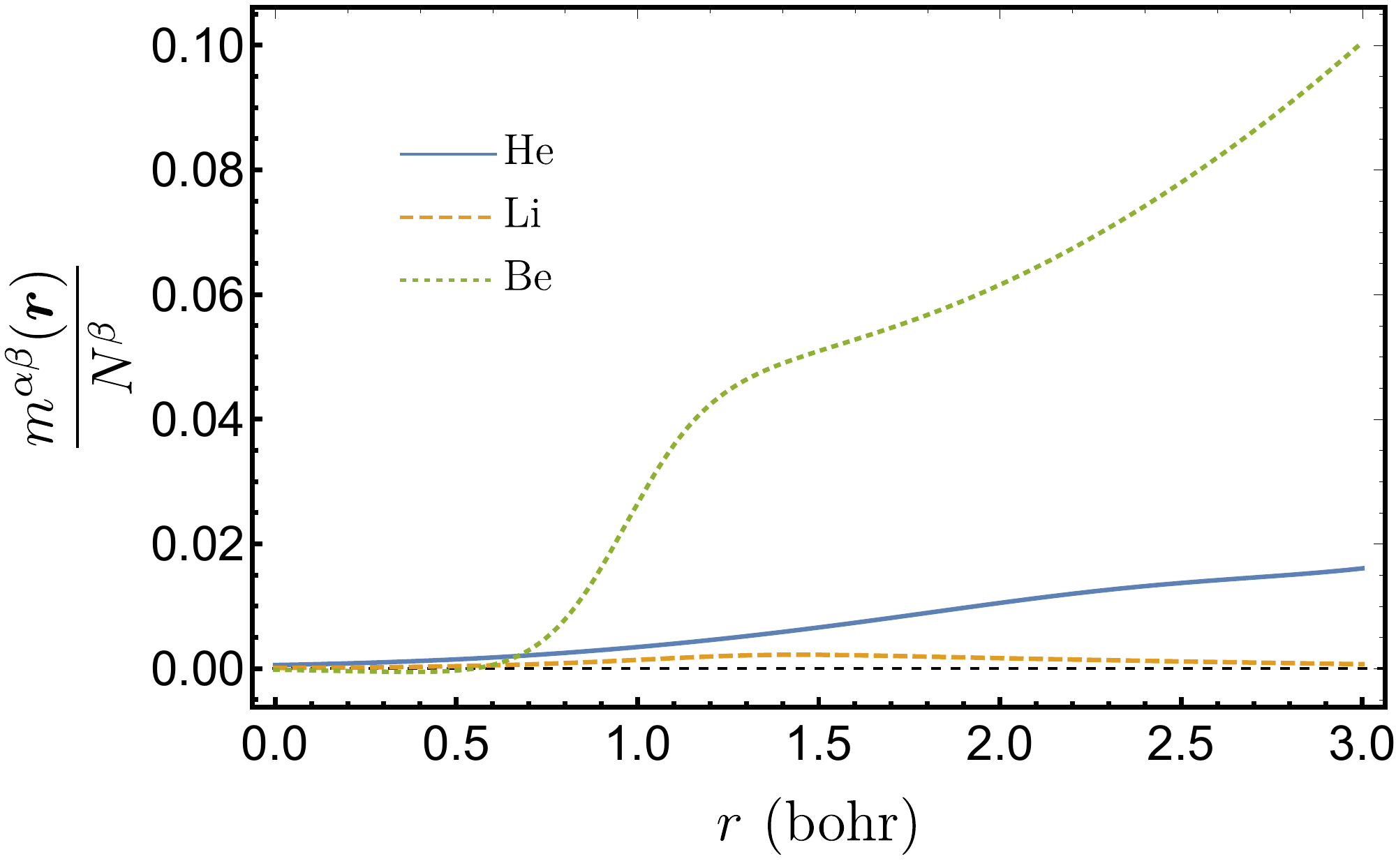} \includegraphics[width=0.5\textwidth]{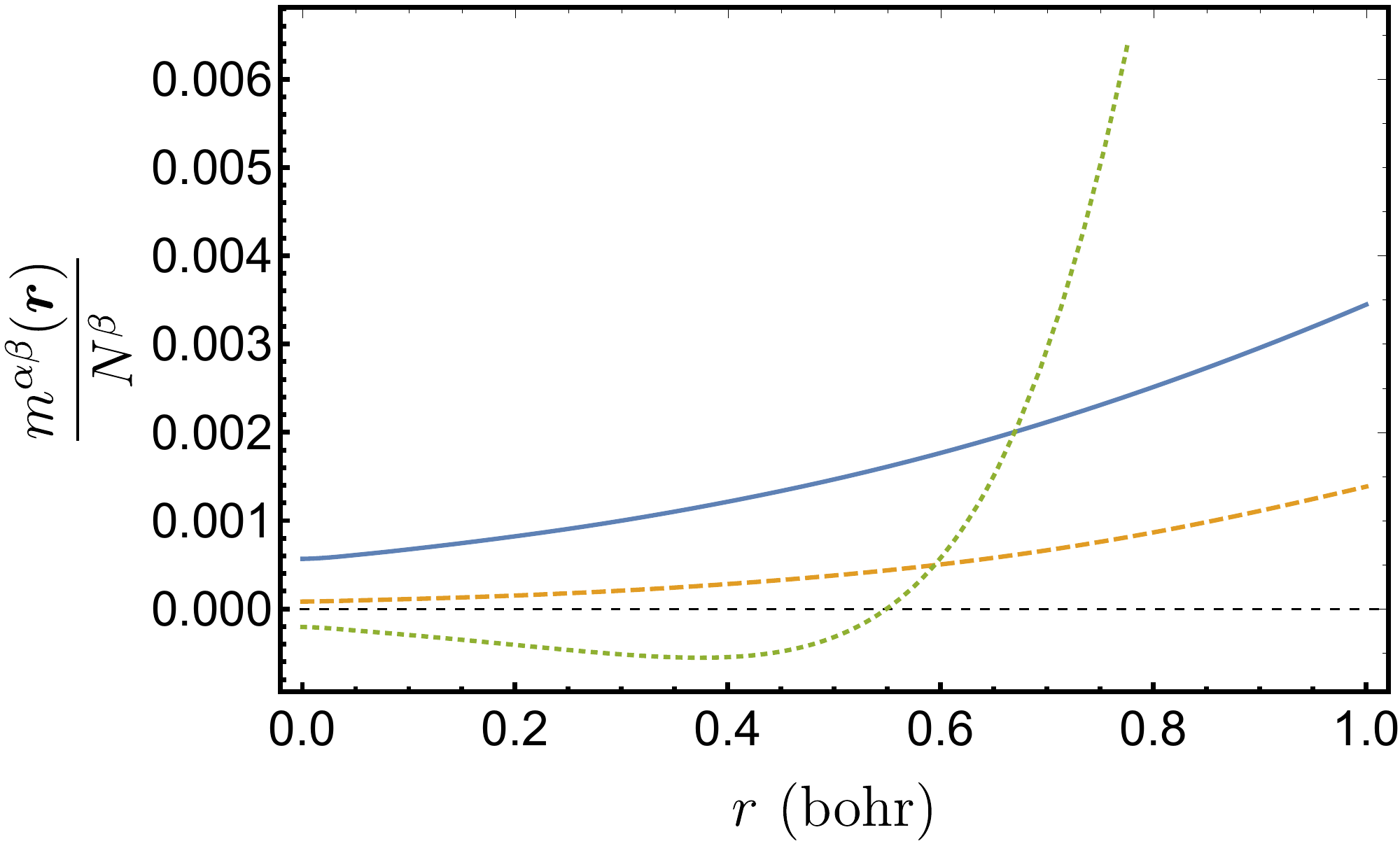}
	\caption{\label{fig:HetoBerMBD} Renormalized $m^{\alpha\beta}(\br)$ of \ce{He}, \ce{Li}, and \ce{Be}.}
	\end{center}
\end{figure}

The reduced MB densities, divided by the number of $\beta$ electrons, of \ce{Be}, \ce{Li} and \ce{He} are presented in Figure \ref{fig:HetoBerMBD}. Note, that for the spin-decomposed reduced MB density, $m^{\sigma\sigma'}(\br) = \frac{m^{\sigma\sigma'}(\br)}{\rho^\sigma(\br)}$.
The $m^{\alpha\beta}(\br)$ are similar, in that they have a minimum at the nucleus and are everywhere positive.  However, the $m^{\alpha\beta}(\br)$ of \ce{Li} reaches a maximum at $\sim$1.4 bohr and then decays due to the fact that $\rho^\alpha(\br)$ decays much slower than $\rho^\beta(\br)$, unlike \ce{He} and \ce{Be}. The amplitude of the \ce{Be} $m^{\alpha\beta}(\br)$, which is dominated by the valence electrons, is significantly larger than that of \ce{He} and \ce{Li}. Near the nucleus $m^{\alpha\beta}(\br)$ of \ce{Be} is negative and relatively flat.  It reaches a minimum at 0.374 bohr and has node at 0.549 bohr.  After crossing the node, $m^{\alpha\beta}(\br)$ increases sharply with a shoulder just beyond $r=1$ bohr.  The $m^{\alpha\beta}(\br)$ of \ce{He} also exhibits a shoulder, although it is less pronounced and further from the nucleus ($\sim$ 2.4 bohr).

\begin{figure}
	\begin{center}
	\includegraphics[width=0.5\textwidth]{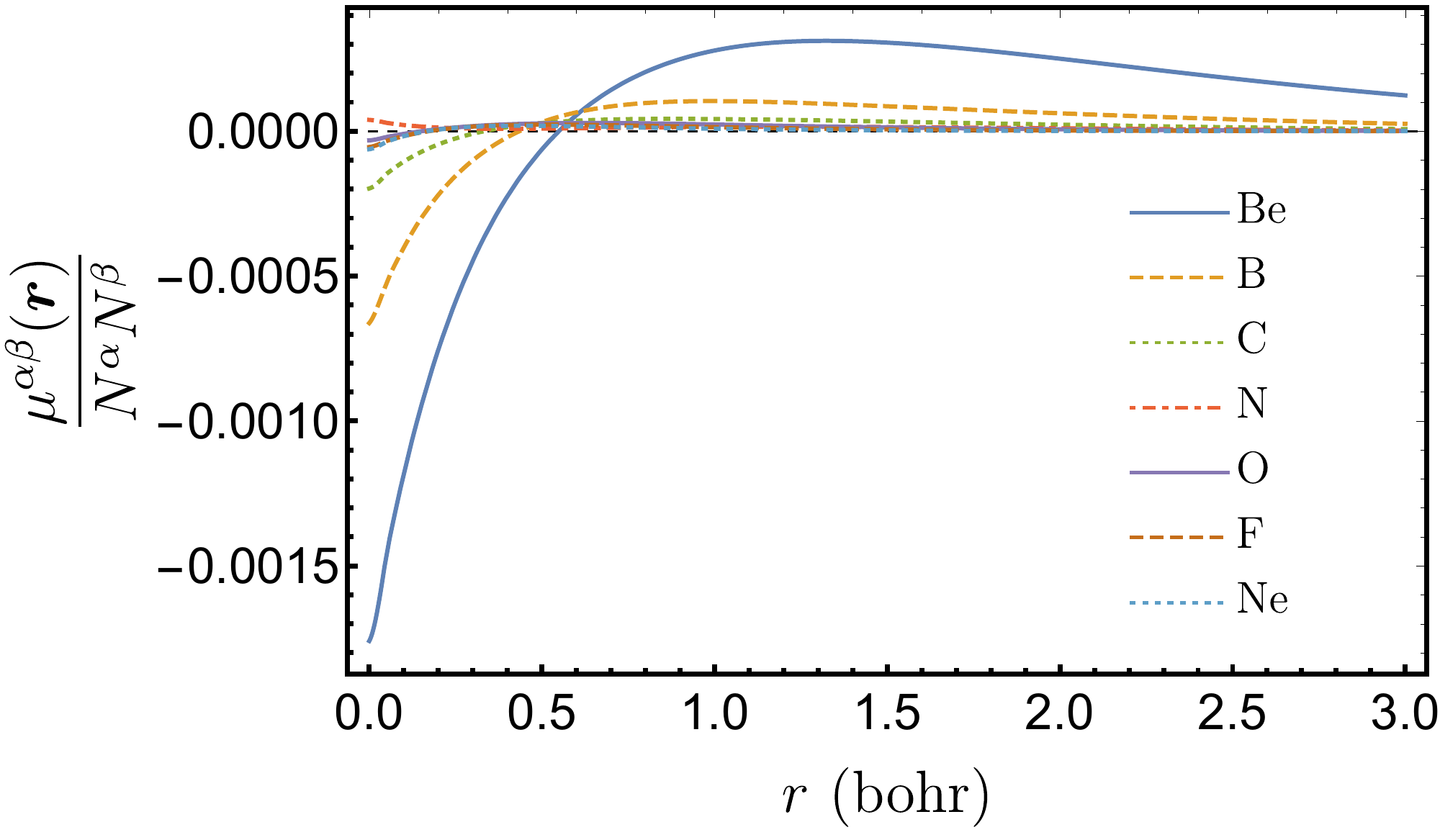} \includegraphics[width=0.5\textwidth]{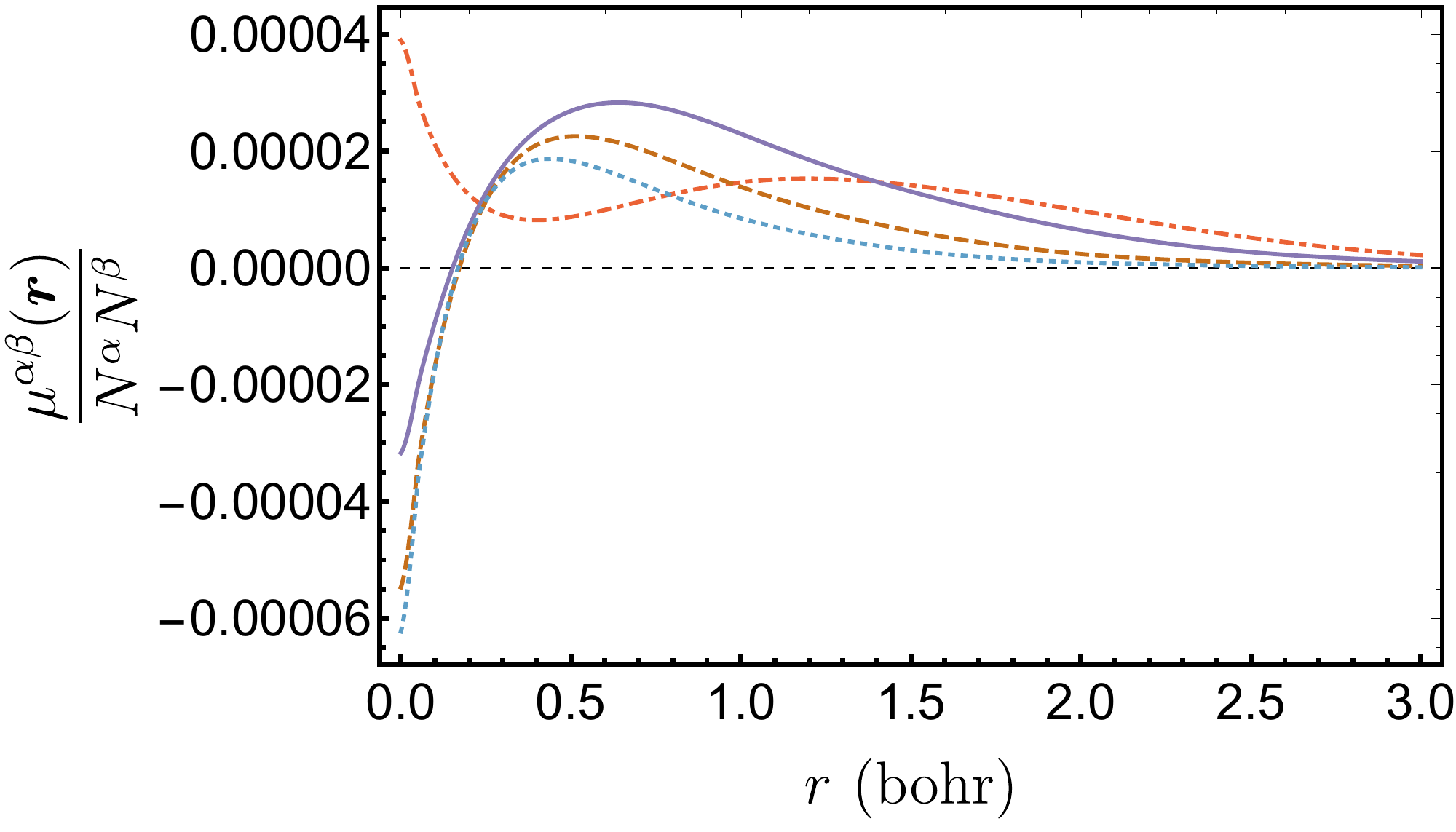}
	\caption{\label{fig:BetoNeMBDab} Renormalized $\mu^{\alpha\beta}(\br)$ of the atoms \ce{Be} to \ce{Ne} (above) and \ce{N} to \ce{Ne} (below).}
	\end{center}
\end{figure}

The $\mu^{\alpha\beta}(\br)$ for \ce{Be} to \ce{Ne} are presented in Figure \ref{fig:BetoNeMBDab}.
With the exception of \ce{N}, all the atoms have a $\mu^{\alpha\beta}(\br)$ with the the same shape as that of \ce{Be}.  Proceeding from \ce{Be} to \ce{C} the amplitude per $\alpha\beta$ electron pair decreases and the node contracts toward the nucleus, in the same manner as the node of a 2$s$ atomic orbital with increasing nuclear charge.  Interestingly, $\mu^{\alpha\beta}(\br)$ for the \ce{N} atom has a positive peak at the nucleus, a minimum near $r = 0.3$ bohr, and is positive everywhere.  This indicates a reduction in the type of correlated motion between the valence electrons, exhibited by the electrons of \ce{Be} to \ce{C}, to such an extent that other correlated motion dominates $\mu^{\alpha\beta}(\br)$, such as the relative motion of core, and core-valence electron pairs.  With the addition of $\beta$-spin $p$-electrons, the $\mu^{\alpha\beta}(\br)$ of \ce{O} to \ce{Ne} have a shape similar to the $\mu^{\alpha\beta}(\br)$ of those atoms before \ce{N}.  From \ce{O} to \ce{Ne}, the amplitude of the maximum per electron pair decreases, while the amplitude of the minimum at the nucleus increases, and the location of the node remains relatively stationary.

\begin{figure}
	\begin{center}
	\includegraphics[width=0.5\textwidth]{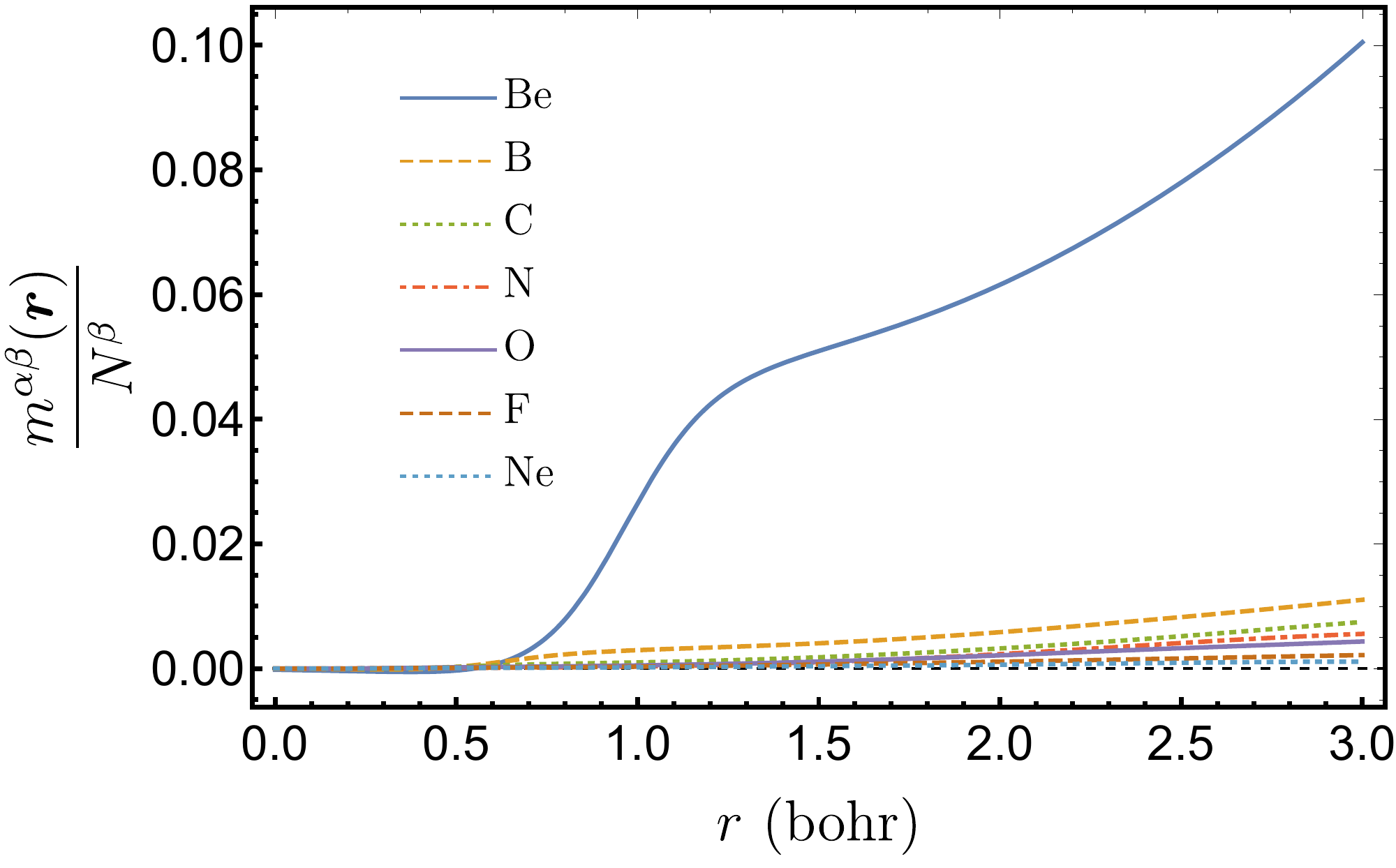} \includegraphics[width=0.5\textwidth]{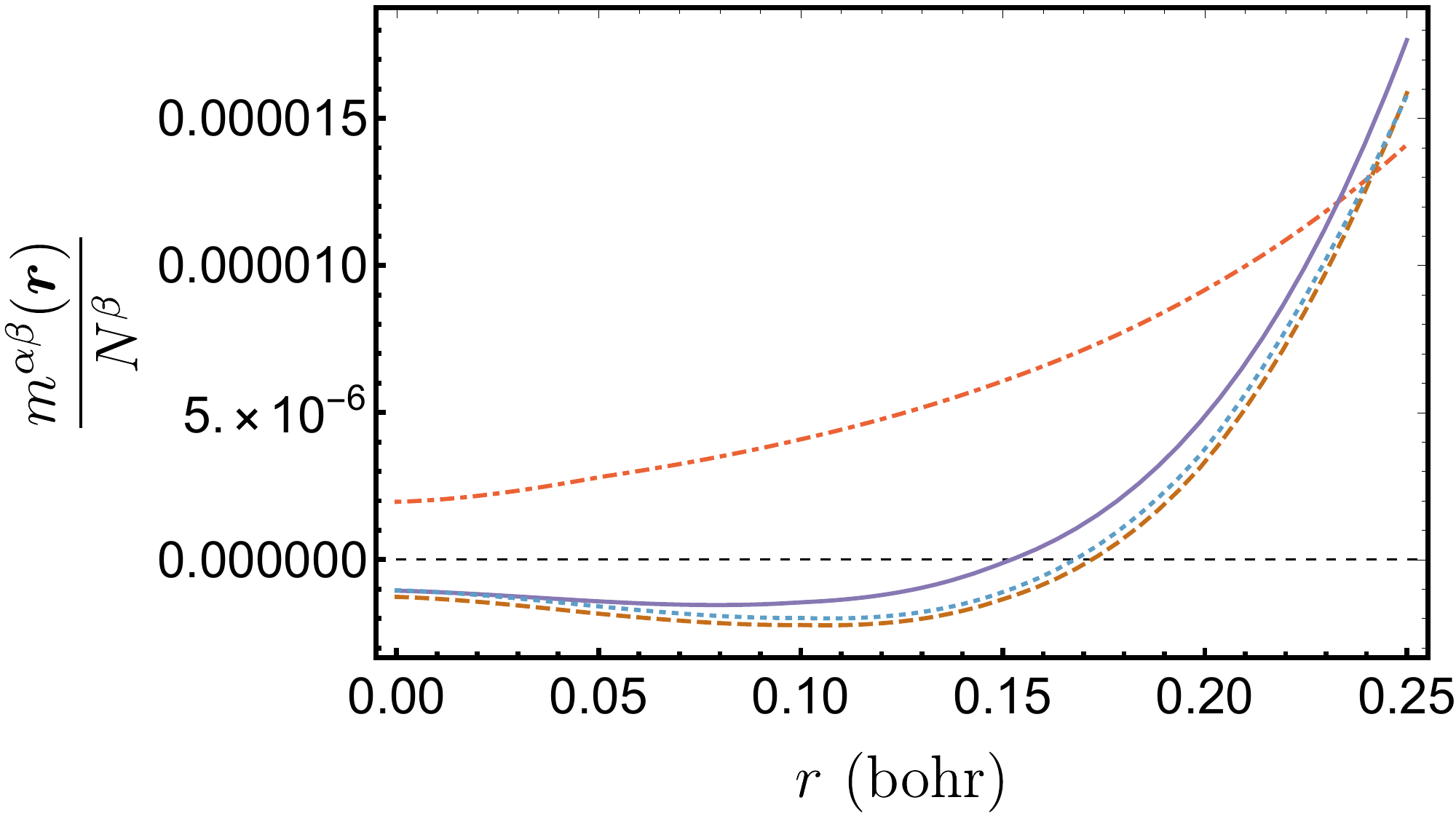}
	\caption{\label{fig:BetoNerMBDab} Renormalized $m^{\alpha\beta}(\br)$ of the atoms \ce{Be} to \ce{Ne} (above) and \ce{N} to \ce{Ne} (below).}
	\end{center}
\end{figure}

For a description of the correlated momemnta that is unbiased by $\rho^\alpha(\br)$, the reduced MB densities, divided by the number of $\beta$ electrons, of \ce{Be} to \ce{Ne} are presented in Figure \ref{fig:BetoNerMBDab}.
As expected from the analysis of $\mu^{\alpha\beta}(\br)$, all $m^{\alpha\beta}(\br)$ have the same shape with the exception of that for \ce{N}, which is positive everywhere.  The amplitude of $m^{\alpha\beta}(\br)/N^\beta$ decreases substantially moving from \ce{Be} to \ce{C}.  Far from the nucleus ($r > 1.5$ bohr), the magnitude of $m^{\alpha\beta}(\br)/N^{\beta}$ also decreases from \ce{O} to \ce{Ne}.  However, the amplitude of $m^{\alpha\beta}(\br)/N^{\beta}$ near the nucleus is less ordered.

\begin{figure}
	\begin{center}
	\includegraphics[width=0.5\textwidth]{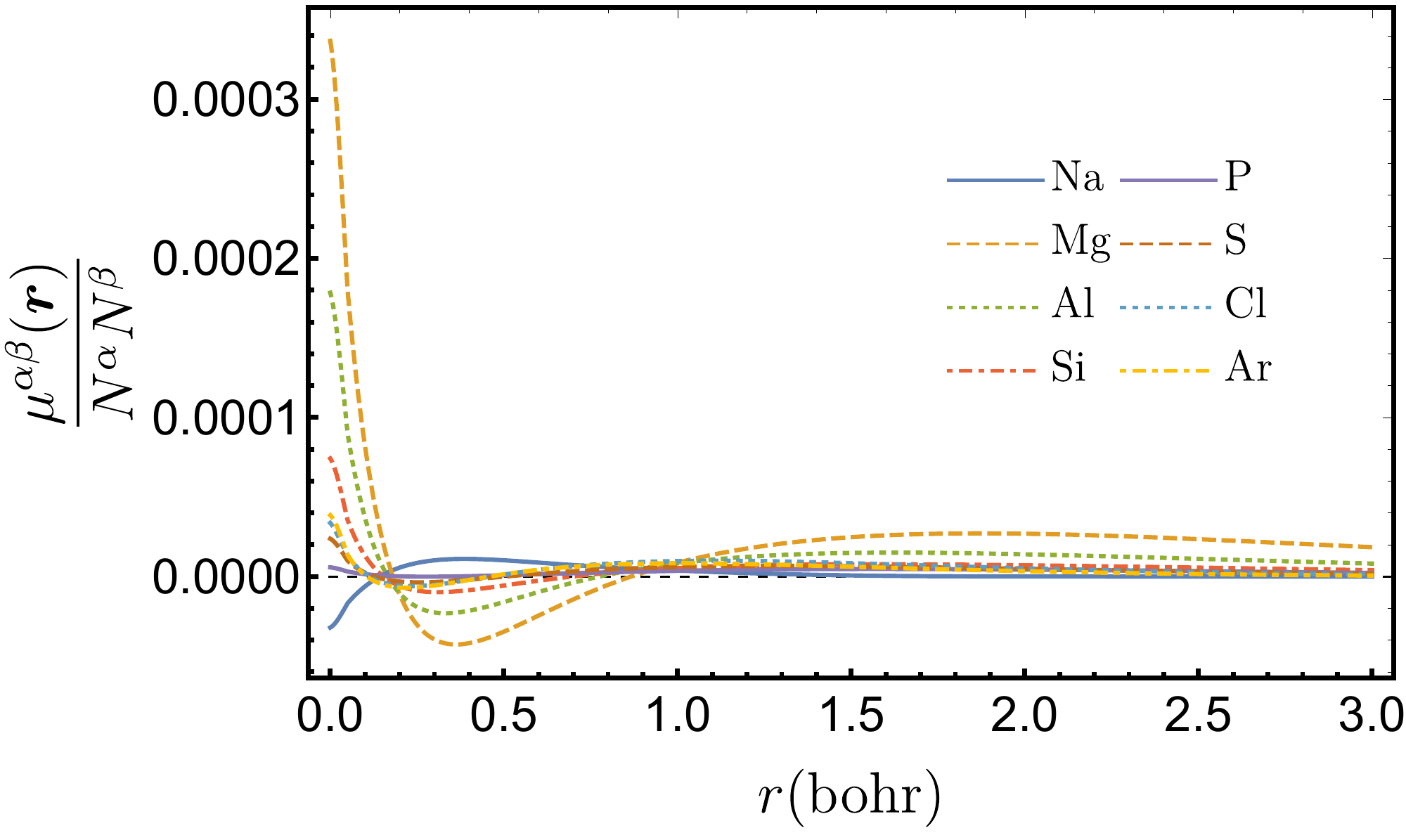}
	\caption{\label{fig:NatoArMBDab} Renormalized $\mu^{\alpha\beta}(\br)$ of the atoms \ce{Na} to \ce{Ar}.}
	\end{center}
\end{figure}

The $\mu^{\alpha\beta}(\br)$, per $\alpha\beta$ electron pair, for the third period atoms are presented in Figure \ref{fig:NatoArMBDab}.
The shape of $\mu^{\alpha\beta}(\br)$ for the \ce{Na} atom is the same as that for \ce{Ne}.  This is due to the fact that the $\beta$-electron configuration remains the same.  Once a $\beta$ electron is added (\ce{Mg} atom and beyond), the magnitude of $\mu^{\alpha\beta}(\br)$ per $\alpha\beta$ electron pair increases substantially, similar to that seen for \ce{Be} in the second period.  Unlike atoms of the second period, the $\mu^{\alpha\beta}(\br)$ of \ce{Mg} and beyond have two nodes (like the $3s$ atomic orbital) and are positive at the nucleus. Yet again, like the second period atoms, $\mu^{\alpha\beta}(\br)>0$ in the valence region and beyond. 

\begin{figure}
	\begin{center}
	\includegraphics[width=0.5\textwidth]{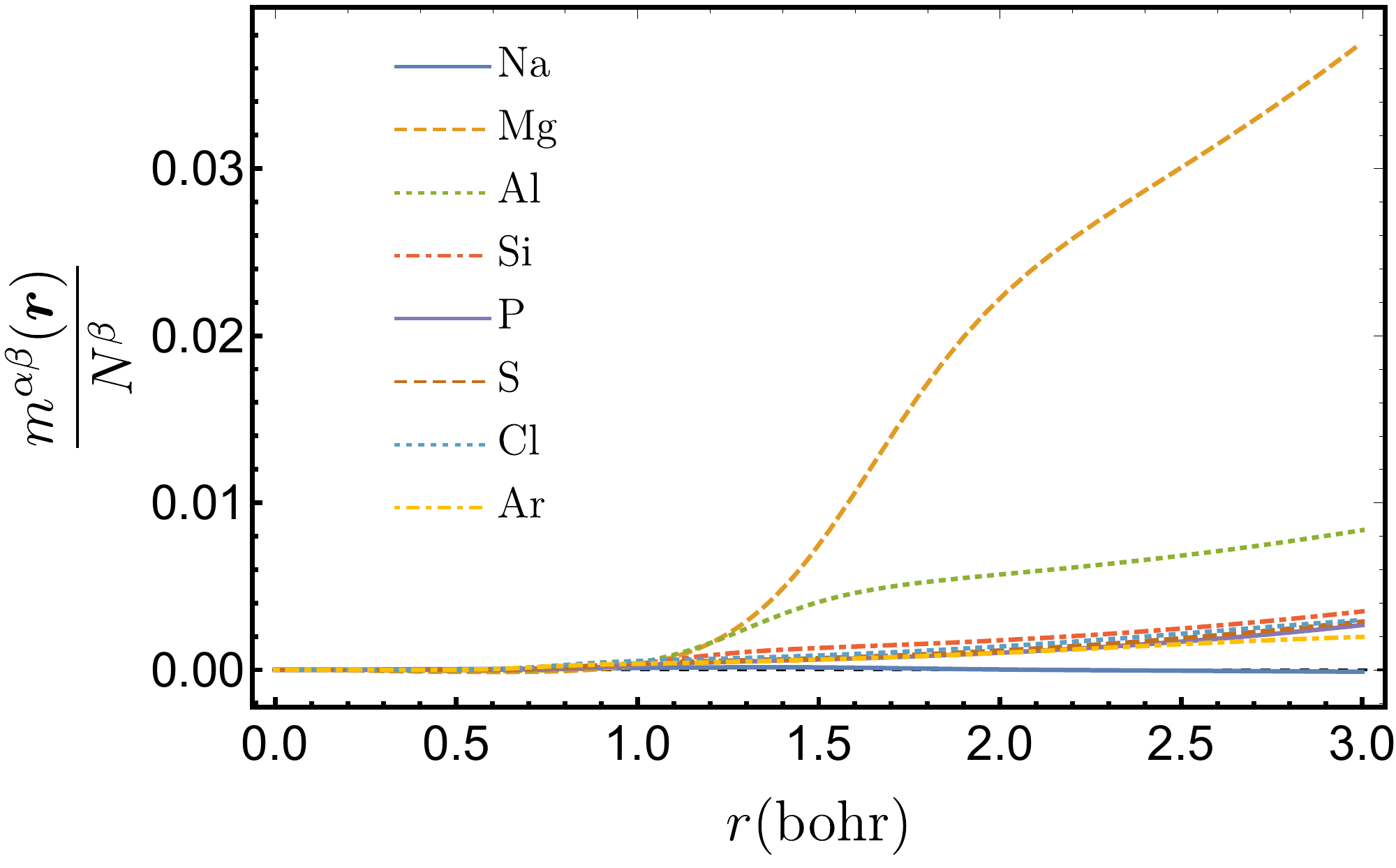} \includegraphics[width=0.5\textwidth]{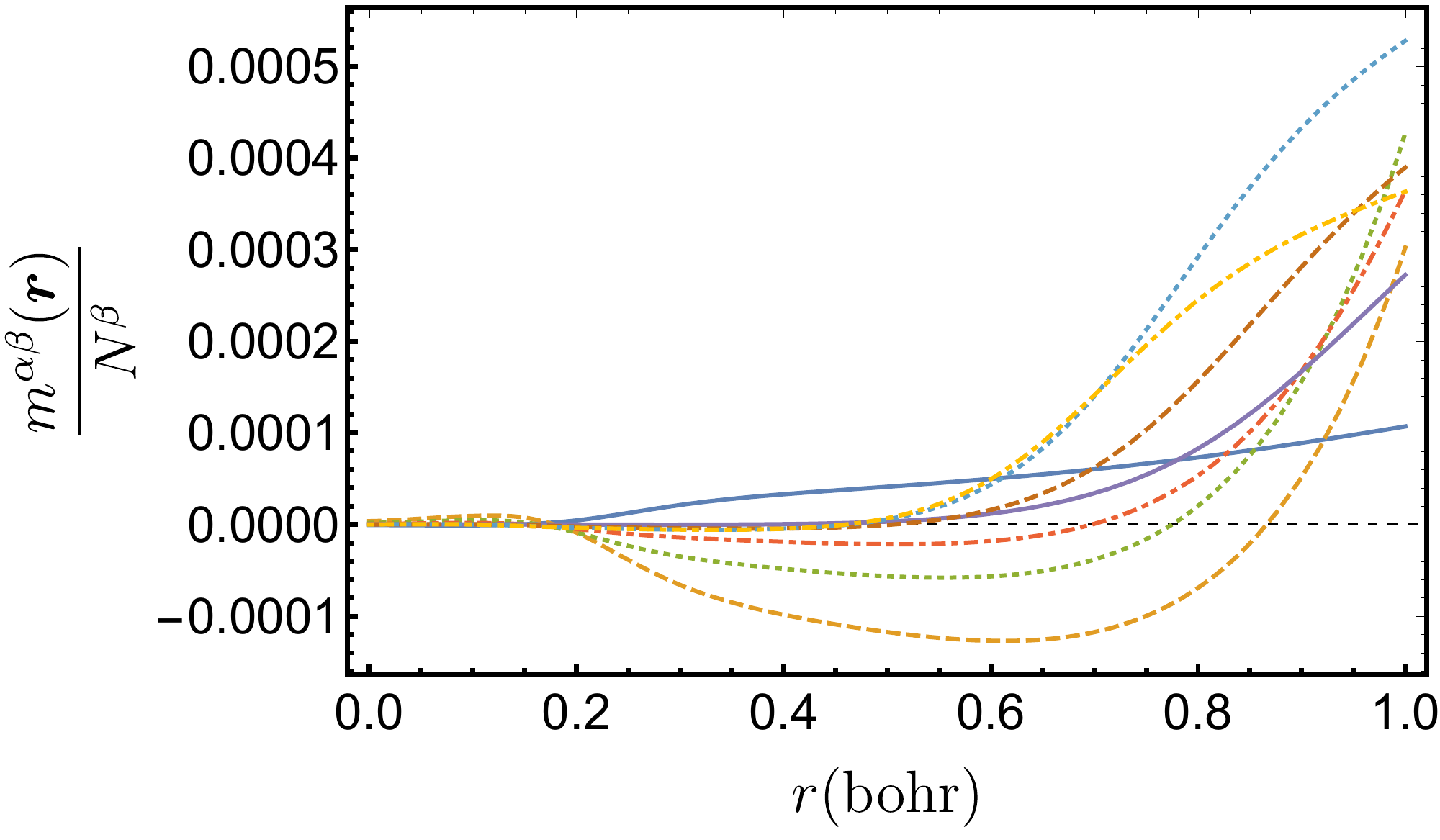} \includegraphics[width=0.5\textwidth]{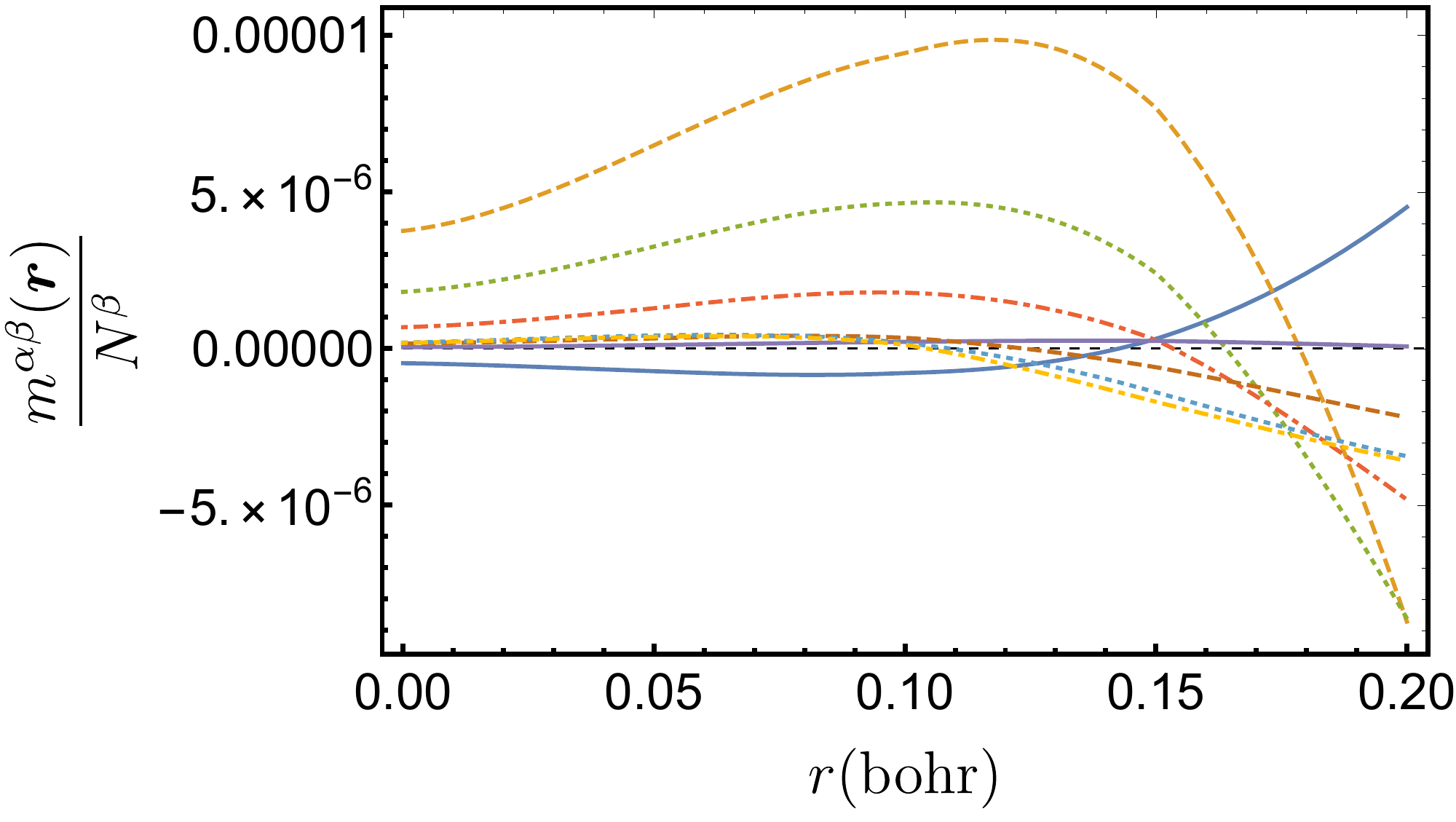}
	\caption{\label{fig:NatoArrMBDab} Renormalized $m^{\alpha\beta}(\br)$ of the atoms \ce{Na} to \ce{Ar}.}
	\end{center}
\end{figure}

The $\alpha\beta$ reduced MB densities, per $\beta$ electron, for the third period atoms are presented in Figure \ref{fig:NatoArrMBDab}.
As suggested by their $\mu^{\alpha\beta}(\br)$, the $m^{\alpha\beta}(\br)$ of \ce{Na} closely resembles that of \ce{Ne}.  The $m^{\alpha\beta}(\br)$ of \ce{Mg}, like \ce{Be} of the second period, has the the largest amplitude per $\beta$ electron. Close to the nucleus, the features of $m^{\alpha\beta}(\br)$ are the same as those for the second period atoms, except the sign is reversed.  These features are nested inside the same features (a minimum and a node) repeated, but with the same sign as those of the second period atoms.  It is also interesting to note that, unlike \ce{N}, the $m^{\alpha\beta}(\br)$ for the $\ce{P}$ atom has the same number of nodes as the other atoms of the period.  However, like \ce{N}, the amplitude of $m^{\alpha\beta}(\br)$ for \ce{P} is relatively minuscule.

\begin{figure}
	\begin{center}
	\includegraphics[width=0.5\textwidth]{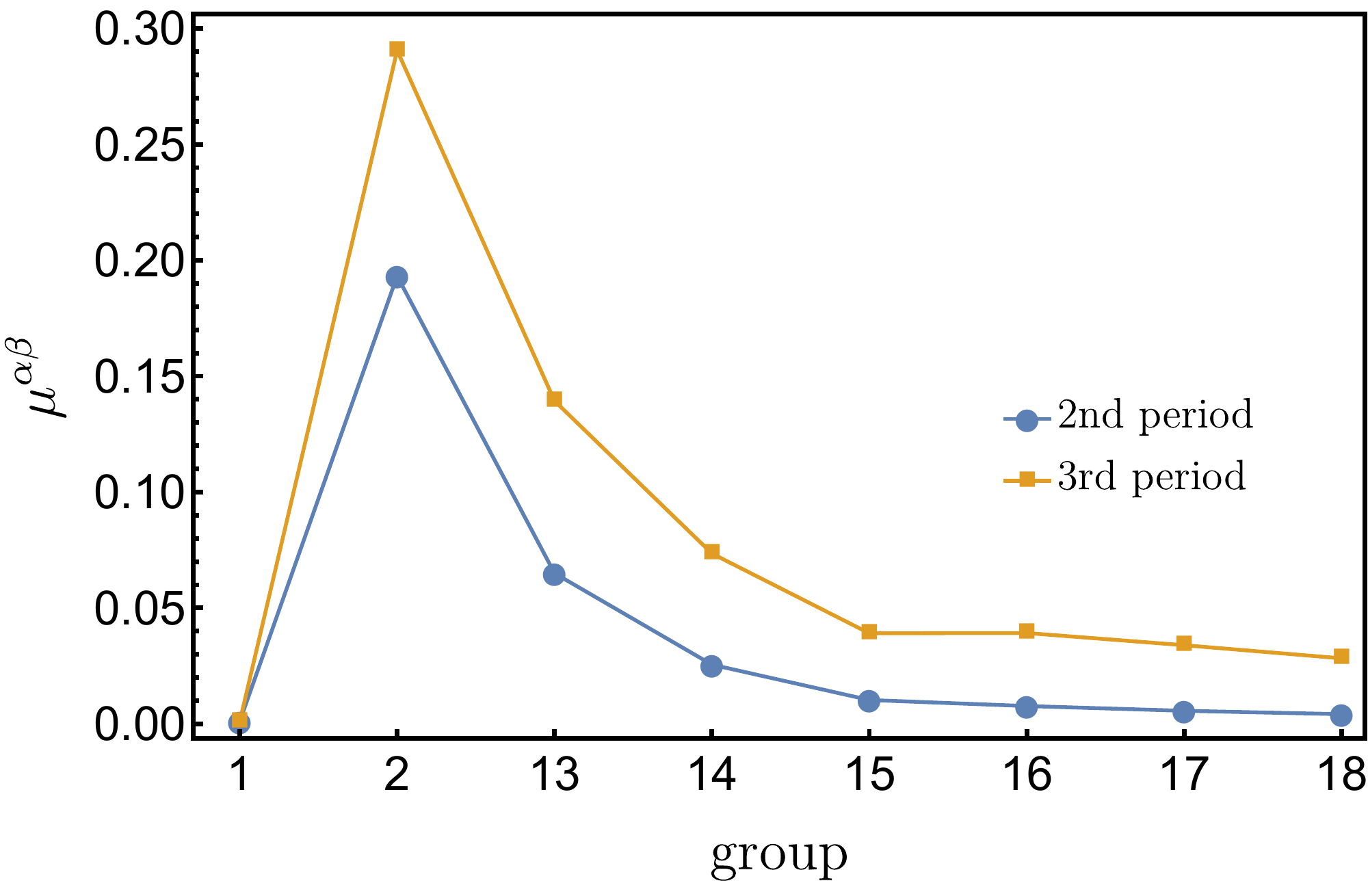}
	\caption{\label{fig:MBab} Opposite-spin momentum-balance of 2nd and 3rd row atoms.}
	\end{center}
\end{figure}

The trend in $\mu^{\alpha\beta}$ for the second and third period atoms is illustrated in Figure \ref{fig:MBab}.
For all atoms $\mu^{\alpha\beta}>0$, which corresponds to opposite-spin electrons preferring to have the same momentum rather than opposite, and periodicity is followed quite closely.  There is a sharp increase in $\mu^{\alpha\beta}$ moving from group 1 to 2 ({\it i.e.} one valence electron to two) which illustrates the importance of valence electron motion to $\mu^{\alpha\beta}$. As atomic number increases, and more electrons are added to the valence shell, the value of $\mu^{\alpha\beta}$ decreases, which suggests less correlated motion.

For open-shell systems $\mu^{\alpha\beta}(\br) \ne \mu^{\beta\alpha}(\br)$, and subsequently $m^{\alpha\beta}(\br) \ne m^{\beta\alpha}(\br)$.  However, the features observed in $\alpha\beta$ densities are very similar to those observed in the $\beta\alpha$ densities.  Therefore, plots of $\mu^{\beta\alpha}(\br)$ and $m^{\beta\alpha}(\br)$ can be found in the supplementary material.

%-----------------------------------
\subsection{Parallel-spin correlation}
%-----------------------------------
\label{subsec:parspin}

As mentioned previously, Fermi correlation contributes to the MB of parallel-spin electrons. Not surprisingly, the contribution of Fermi correlation is orders of magnitude larger than that of Coulomb correlation.  Consider the \ce{Ne} atom, where $\mu_\text{HF}^{\sigma\sigma}(\br) = -0.08644$ and $\mu^{\sigma\sigma}(\br) = -0.08663$, which means Coulomb correlation contributes $\sim$0.2\% to the overall parallel-spin MB.  Furthermore, Fermi correlation favours $\mu^{\sigma\sigma}<0$, which is parallel-spin electrons having the exact opposite momentum.

\begin{figure}
	\begin{center}
	\includegraphics[width=0.5\textwidth]{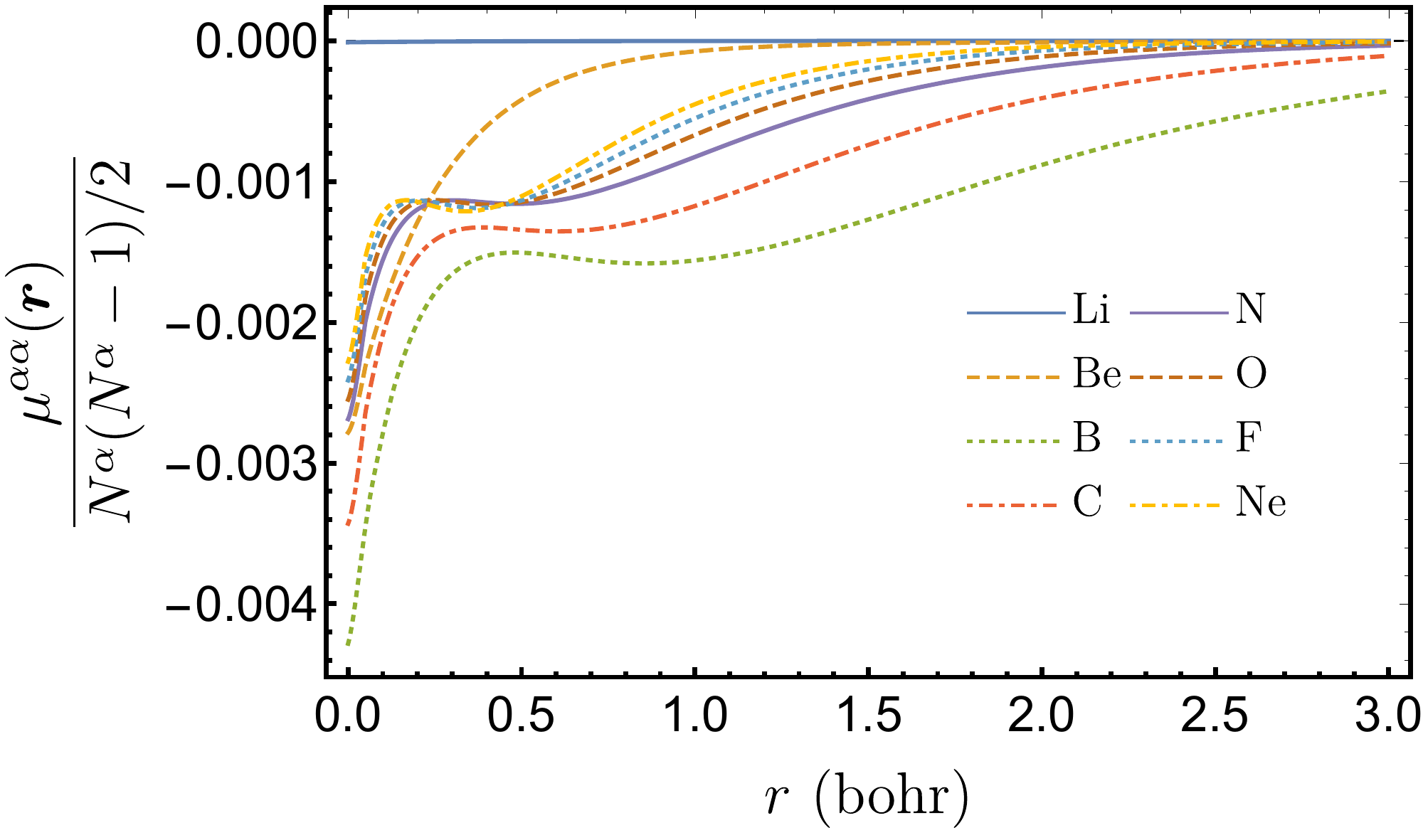}
	\caption{\label{fig:BetoNeMBDaa} Renormalized $\mu^{\alpha\alpha}(\br)$ of the atoms \ce{Li} to \ce{Ne}.}
	\end{center}
\end{figure}

The $\alpha\alpha$ MB density per electron pair, $\frac{\mu^{\alpha\alpha}(\br)}{N^{\alpha}(N^{\alpha}-1)/2}$, for atoms of the second period are presented in Figure \ref{fig:BetoNeMBDaa}.
It is seen that $\mu^{\alpha\alpha}(\br)$ is negative everywhere, meaning all regions contribute to electrons having the exact opposite momentum.  Similar to the opposite-spin MB density, the amplitude of $\mu^{\alpha\alpha}(\br)$ is relatively small for \ce{Li}.  For the \ce{Be} atom, which has only one valence $\alpha$-electron, the amplitude of $\mu^{\alpha\alpha}(\br)$ is much more appreciable. But, it is still significantly smaller than that for atoms with multiple valence $\alpha$-electrons.  The addition of $p$-electrons for \ce{B}, and the proceeding atoms, creates a shoulder in $\mu^{\alpha\alpha}(\br)$ further from the nucleus.  Moving from \ce{B} to \ce{Ne} the shoulder contracts to the nucleus and the amplitude of $\mu^{\alpha\alpha}(\br)$ per electron pair decreases.

\begin{figure}
	\begin{center}
	\includegraphics[width=0.5\textwidth]{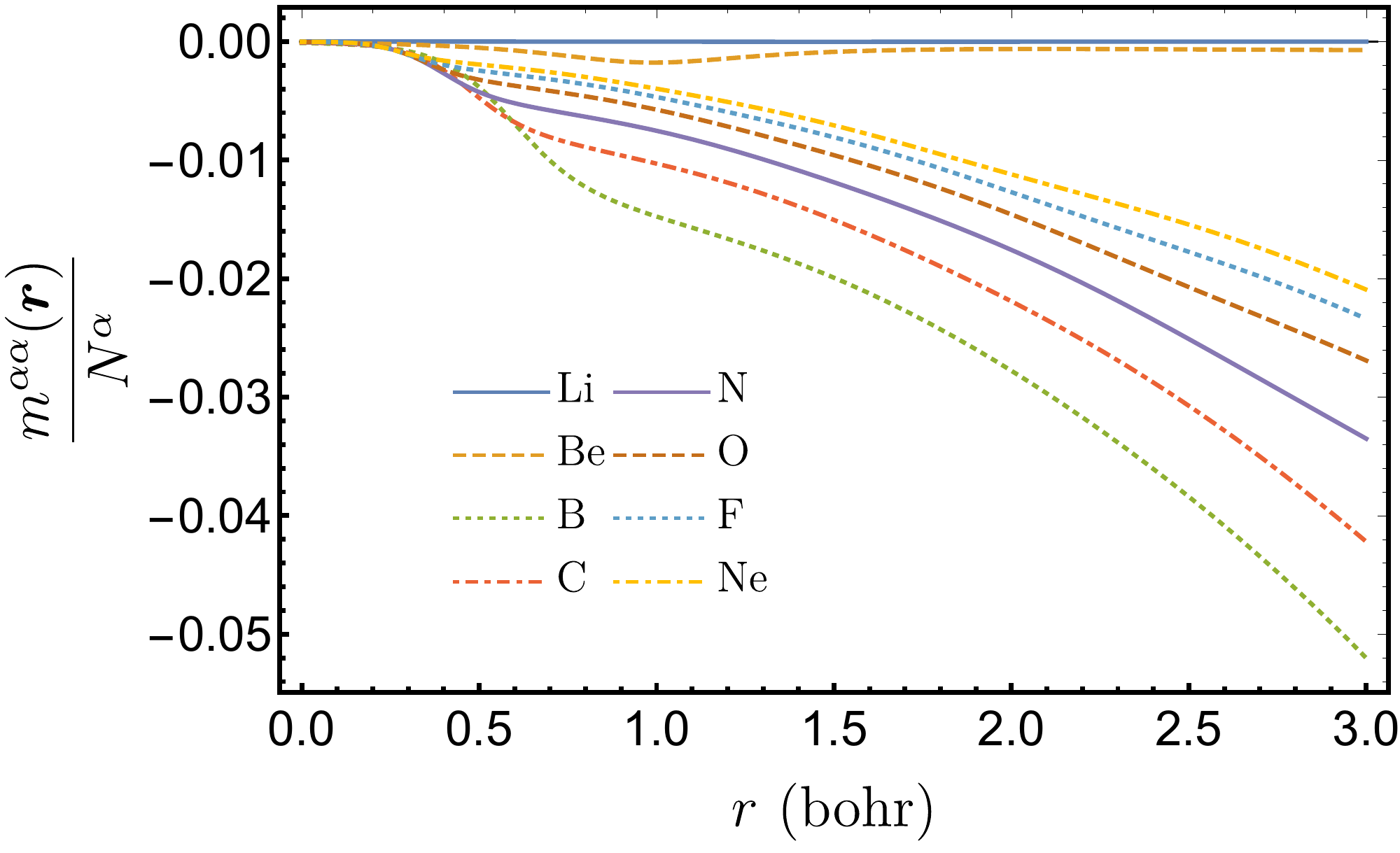}
	\caption{\label{fig:BetoNerMBDaa} Renormalized $m^{\alpha\alpha}(\br)$ of the atoms \ce{Li} to \ce{Ne}.}
	\end{center}
\end{figure}

Figure \ref{fig:BetoNerMBDaa} presents the $\alpha\alpha$ reduced MB density, divided by the number of $\alpha$ electrons, for the second period atoms.
The amplitude of $m^{\alpha\alpha}(\br)/N^\alpha$ increases slightly from \ce{Li} to \ce{Be}, and then substantially from \ce{Be} to \ce{B}. This is accompanied by a change in the shape of $m^{\alpha\alpha}(\br)$ due to the addition of a $p$-electron.  The amplitude for \ce{B} is the largest amongst the atoms of the second period, and the amplitude gradually decreases from \ce{B} to \ce{Ne}.  As with opposite-spin MB density, the magnitude of the reduced MB is a minimum at the nucleus, and the $m^{\alpha\alpha}(\br)/N^\alpha$ converge to a small range of values at the origin. 

\begin{figure}
	\begin{center}
	\includegraphics[width=0.5\textwidth]{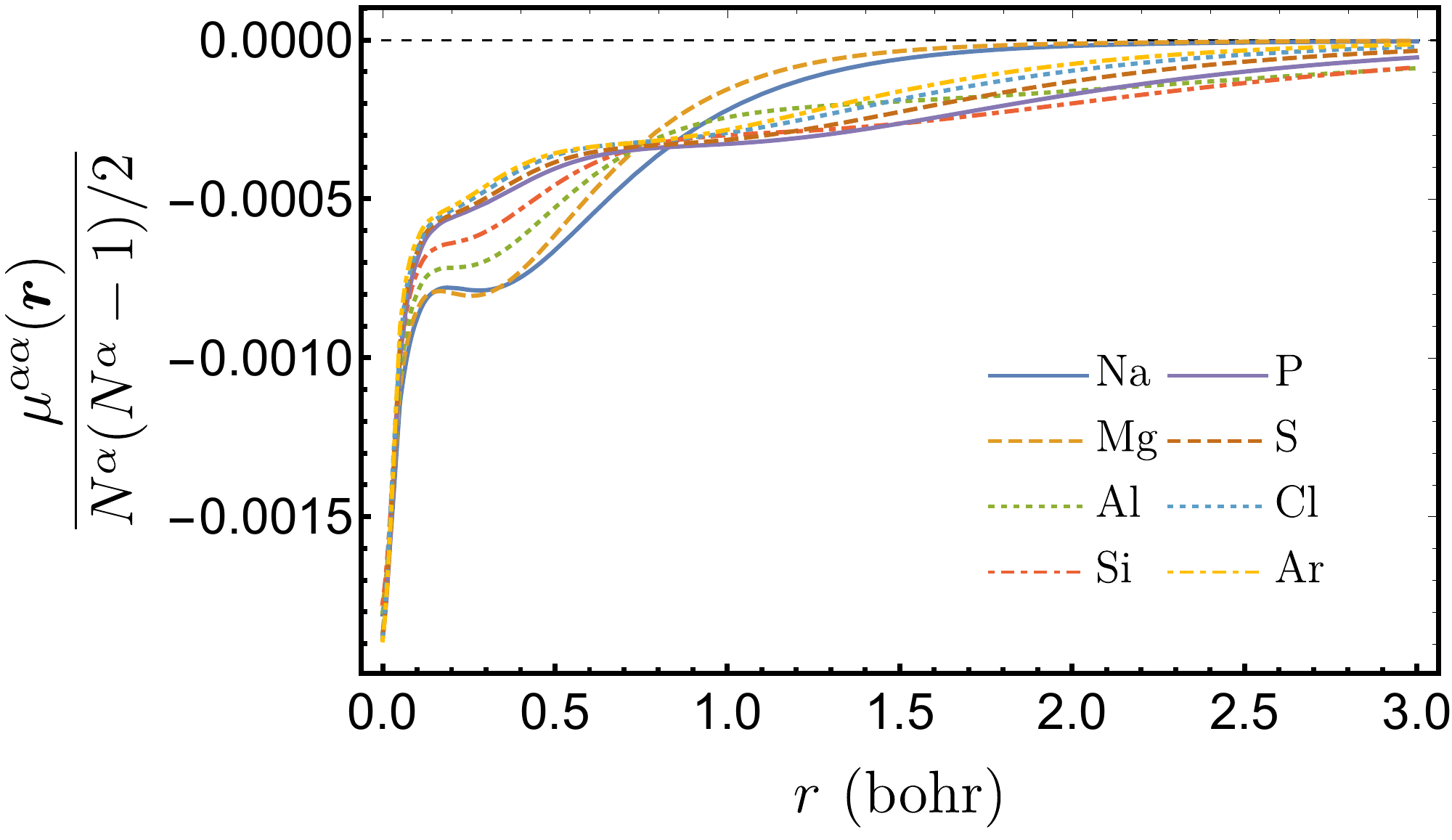}
	\caption{\label{fig:NatoArMBDaa} Renormalized $\mu^{\alpha\alpha}(\br)$ of the atoms \ce{Na} to \ce{Ar}.}
	\end{center}
\end{figure}

The $\mu^{\alpha\alpha}(\br)$ per electron pair for the third period atoms are plotted in Figure \ref{fig:NatoArMBDaa}.
Similar to the $\mu^{\alpha\alpha}(\br)$ of the second period atoms, the $\mu^{\alpha\alpha}(\br)$ of the third period are negative everywhere.  The $\mu^{\alpha\alpha}(\br)$ of the third period atoms have extra features due to the presence of additional core electrons.  The $\mu^{\alpha\alpha}(\br)$ of \ce{Na} and \ce{Mg} resemble that of \ce{Ne}.  The secondary minimum near the nucleus decreases in magnitude and almost disappears as $\alpha$ $p$-electrons are added from \ce{Al} to \ce{P}.  For those atoms with $\alpha$ $p$-electrons, there is a less pronounced shoulder at larger $r$.

\begin{figure}
	\begin{center}
	\includegraphics[width=0.5\textwidth]{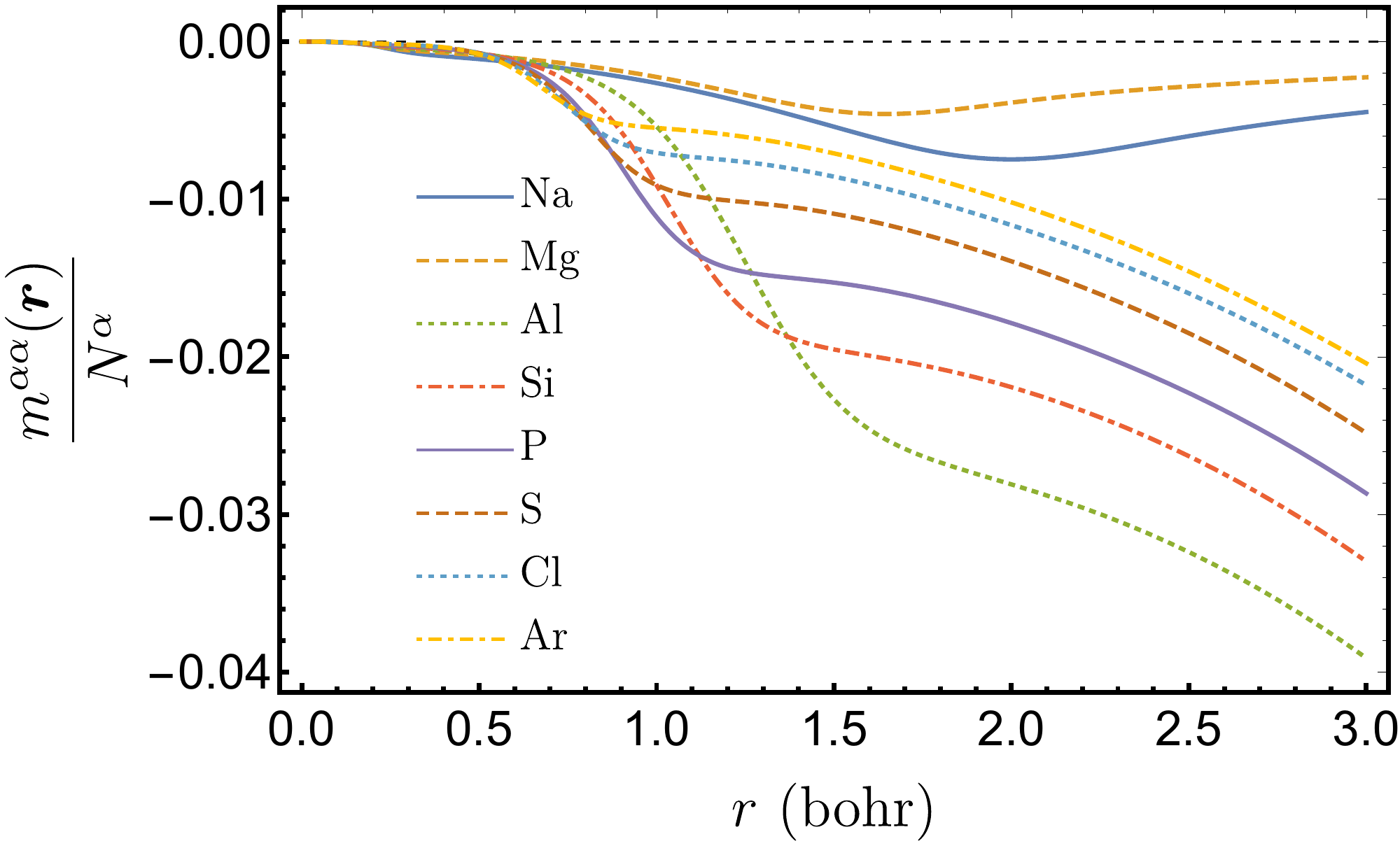} \includegraphics[width=0.5\textwidth]{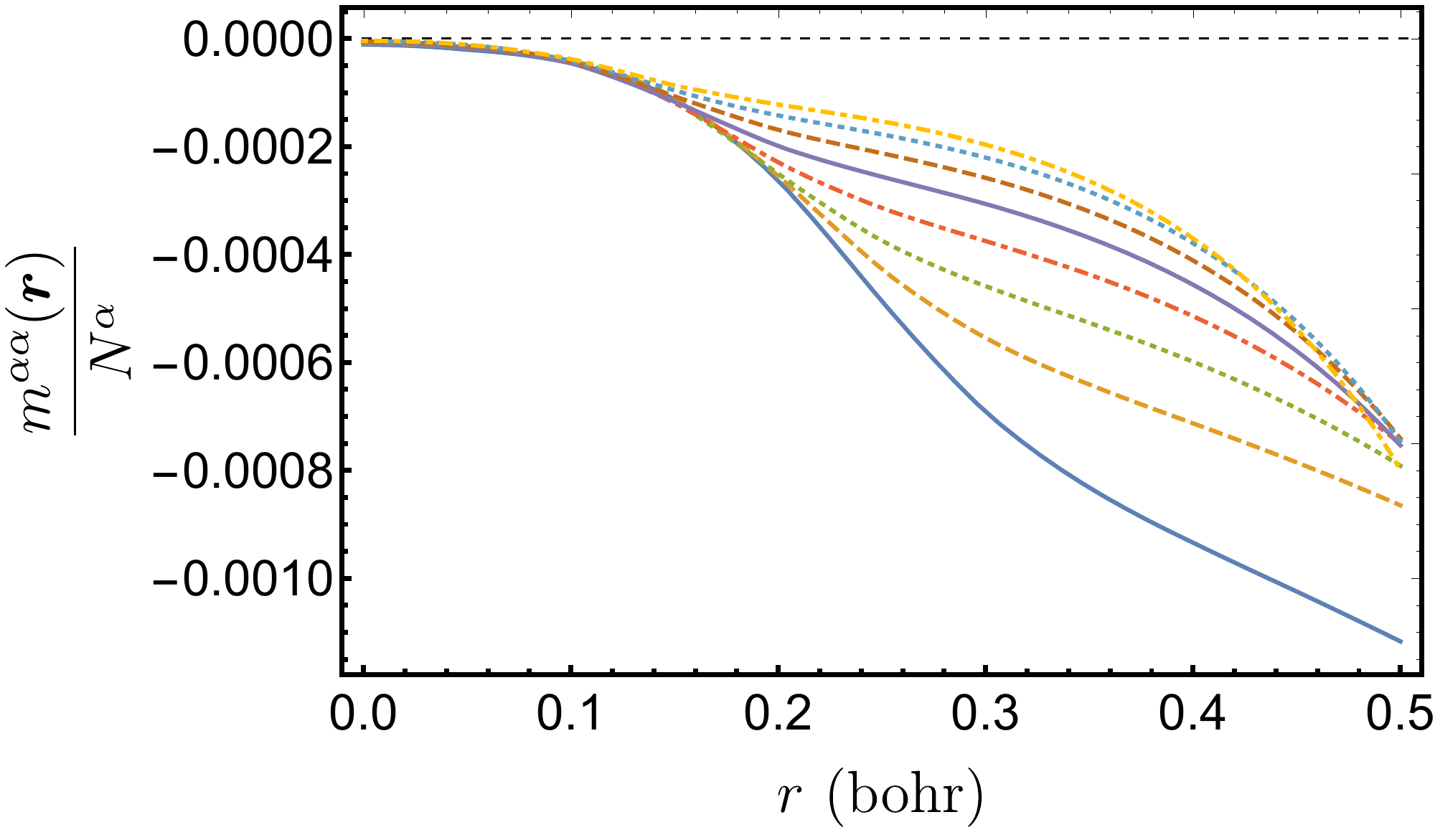}
	\caption{\label{fig:NatoArrMBDaa} Renormalized $m^{\alpha\alpha}(\br)$ of the atoms \ce{Na} to \ce{Ar}.}
	\end{center}
\end{figure}

The $\alpha\alpha$ reduced MB density per $\alpha$ electron is presented for the third period atoms in Figure \ref{fig:NatoArrMBDaa}. 
The transition from atoms without valence $p$-electrons to those with them is obvious.  The $m^{\alpha\alpha}(\br)$ of both \ce{Na} and \ce{Mg} have minima in their valence region (similar to \ce{Li} and \ce{Be}), whereas the atoms beyond \ce{Mg} exhibit a shoulder and continue to increase in amplitude.  Also similar to the second period, the maximum amplitude of $m^{\alpha\alpha}(\br)/N^{\alpha}$ is reached by the atom with only one $p$-electron and the amplitude decreases across the period.  Furthermore, the values of $m^{\alpha\alpha}(\br)/N^{\alpha}$ converge to a short range of near-zero values close to the nucleus.

Finally, the trend in the total MB between parallel-spin electrons for atoms of the second and third period is illustrated in Figure \ref{fig:MBaabb}. The actual values can be found in Table I of the supplementary material.
\begin{figure}
	\begin{center}
	\includegraphics[width=0.5\textwidth]{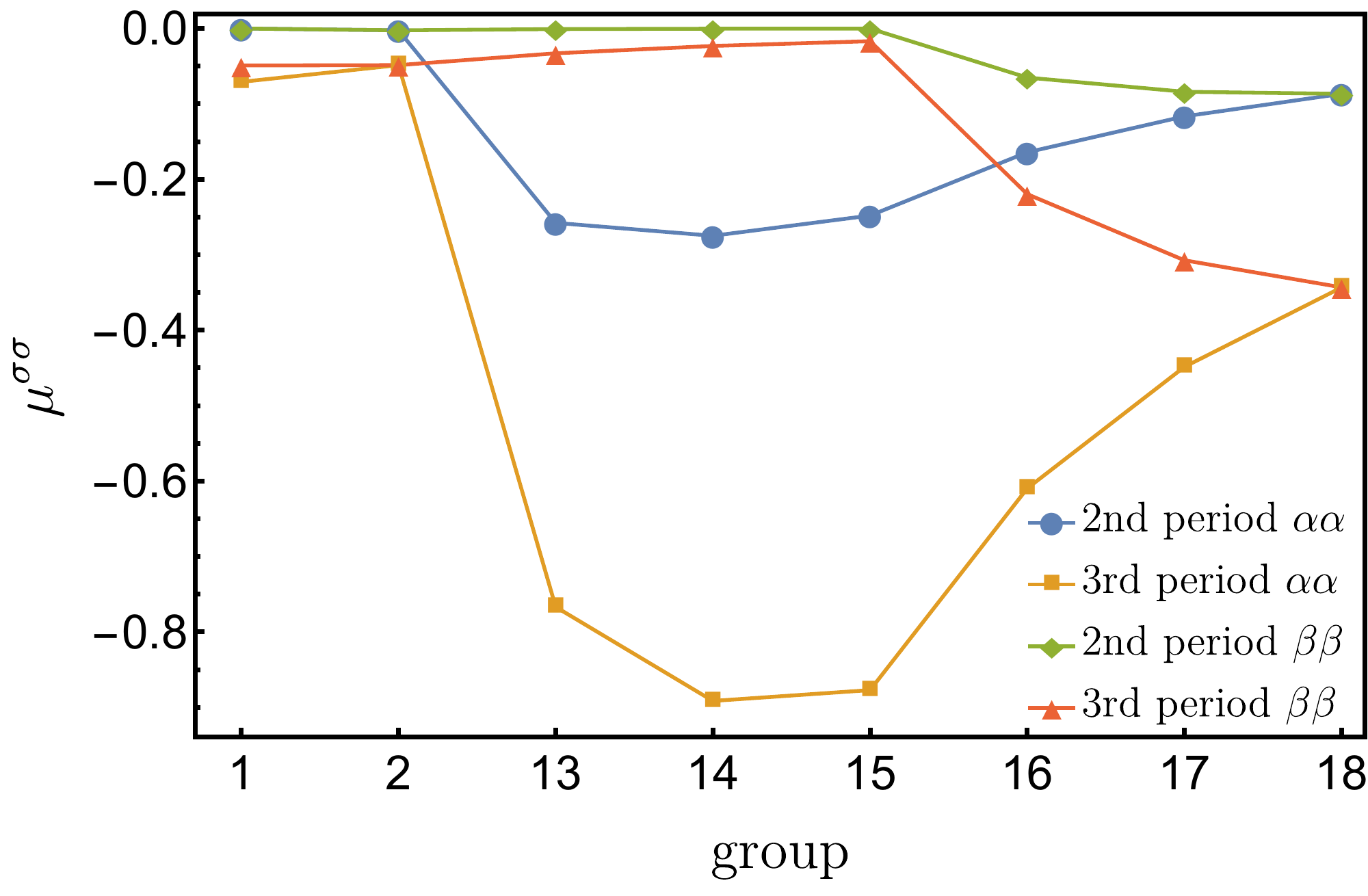}
	\caption{\label{fig:MBaabb} Parallel-spin momentum balance of 2nd and 3rd row atoms.}
	\end{center}
\end{figure}
Fermi correlation, or exchange, is mainly responsible for the shape of $\mu^{\sigma\sigma}(\br)$ and the subsequent value of $\mu^{\sigma\sigma}$.  In agreement with the findings of Koga,\cite{Koga2002} concerning the the angular separation of electron momenta in atoms, the magnitude of MB is strongly dependent on the $p$-electrons.  This is seen with the dramatic decrease in the value of $\mu^{\alpha\alpha}$ from group 2 to 13, and the decrease in $\mu^{\beta\beta}$ seen from group 15 to 16.  Interestingly, the minimum value of $\mu^{\alpha\alpha}$ occurs at group 14, rather than group 15 which has the maximum number of unpaired $p$-electrons.  Also, it is seen that $\mu^{\beta\beta}$ slightly decreases when $\alpha$ $p$-electrons are added (group 2 to 15), and $\mu^{\alpha\alpha}$ decreases significantly as $\beta$ $p$-electrons are added (group 15 to 17).

The decrease in correlated electron motion, as indicated by the decrease in the magnitude of $\mu^{\alpha\alpha}$, and the actual overall parallel-spin MB ($\mu^{\alpha\alpha} + \mu^{\beta\beta}$), from group 14 to 18, is similar to the decrease seen in opposite-spin MB as the number of valence electrons is increased (\ref{fig:MBab}). Due to the calculation of results short of the complete basis set limit, it is somewhat precarious to draw firm conclusions regarding the nature of this decrease for opposite-spin electrons alone.  However, $>$99\% of parallel-spin MB is due to Fermi correlation which, when considering magnitude, has been modelled precisely.  Therefore, the most reasonable explanation for the reduction in correlated motion amongst parallel-spin electrons is the decrease in space available as the electrons are crowded into the valence region.  It is expected that the same phenomenon affects the Coulomb correlation of opposite-spin electrons. 

As with the opposite-spin MB density analysis, the parallel MB densities and reduced MB densities for $\beta$ electrons are given in the supplementary material due to the similarities in features and trends compared to the $\alpha$ electron densities.

%-----------------------------------
\subsection{Comparing correlation}
\begin{figure}
	\begin{center}
	\includegraphics[width=0.5\textwidth]{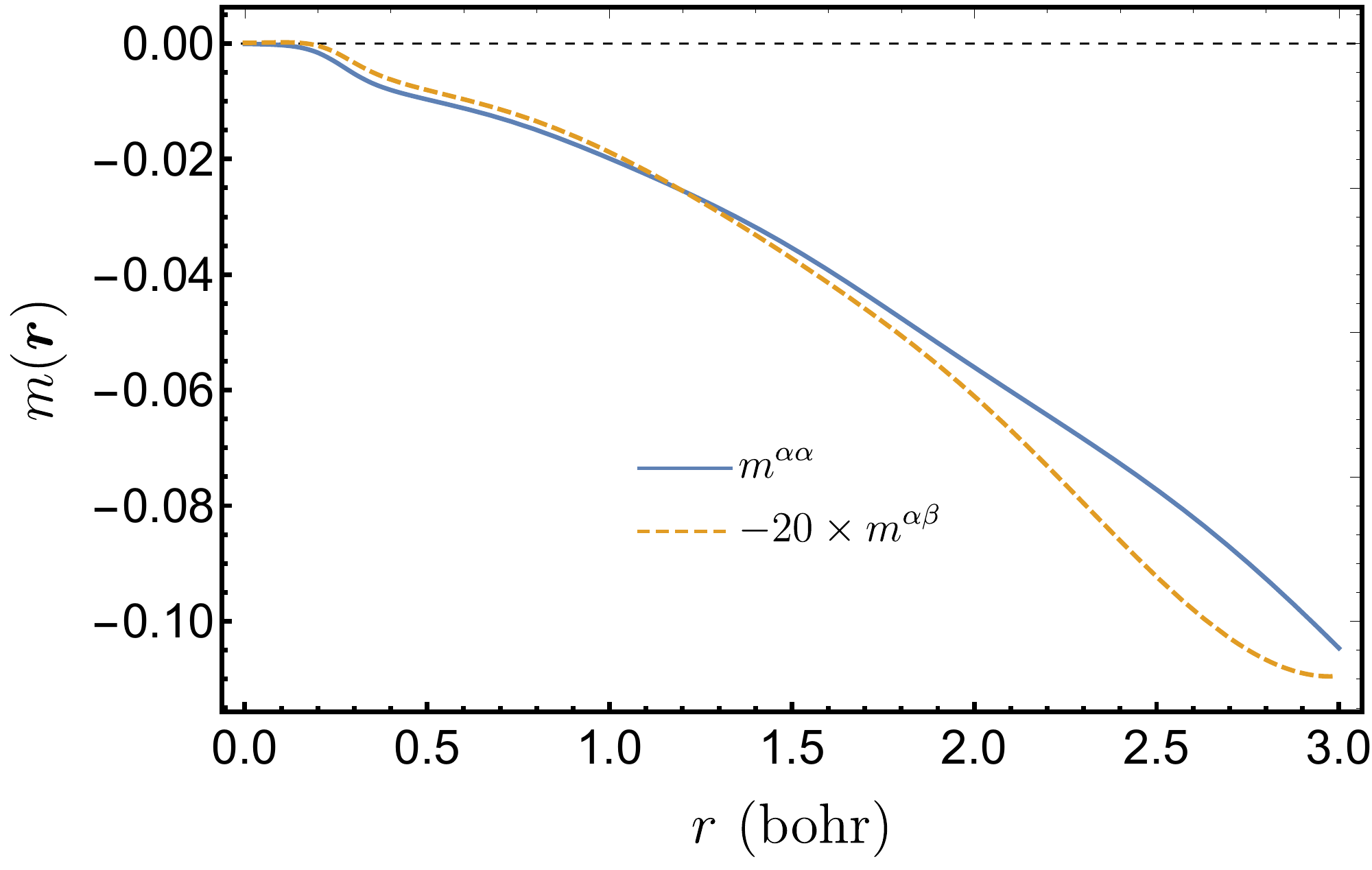}
	\caption{\label{fig:NerMBD} Comparison between $m^{\alpha\alpha}(\br)$ and scaled $m^{\alpha\beta}(\br)$ of \ce{Ne}.}
	\end{center}
\end{figure}

\begin{figure}
	\begin{center}
	\includegraphics[width=0.5\textwidth]{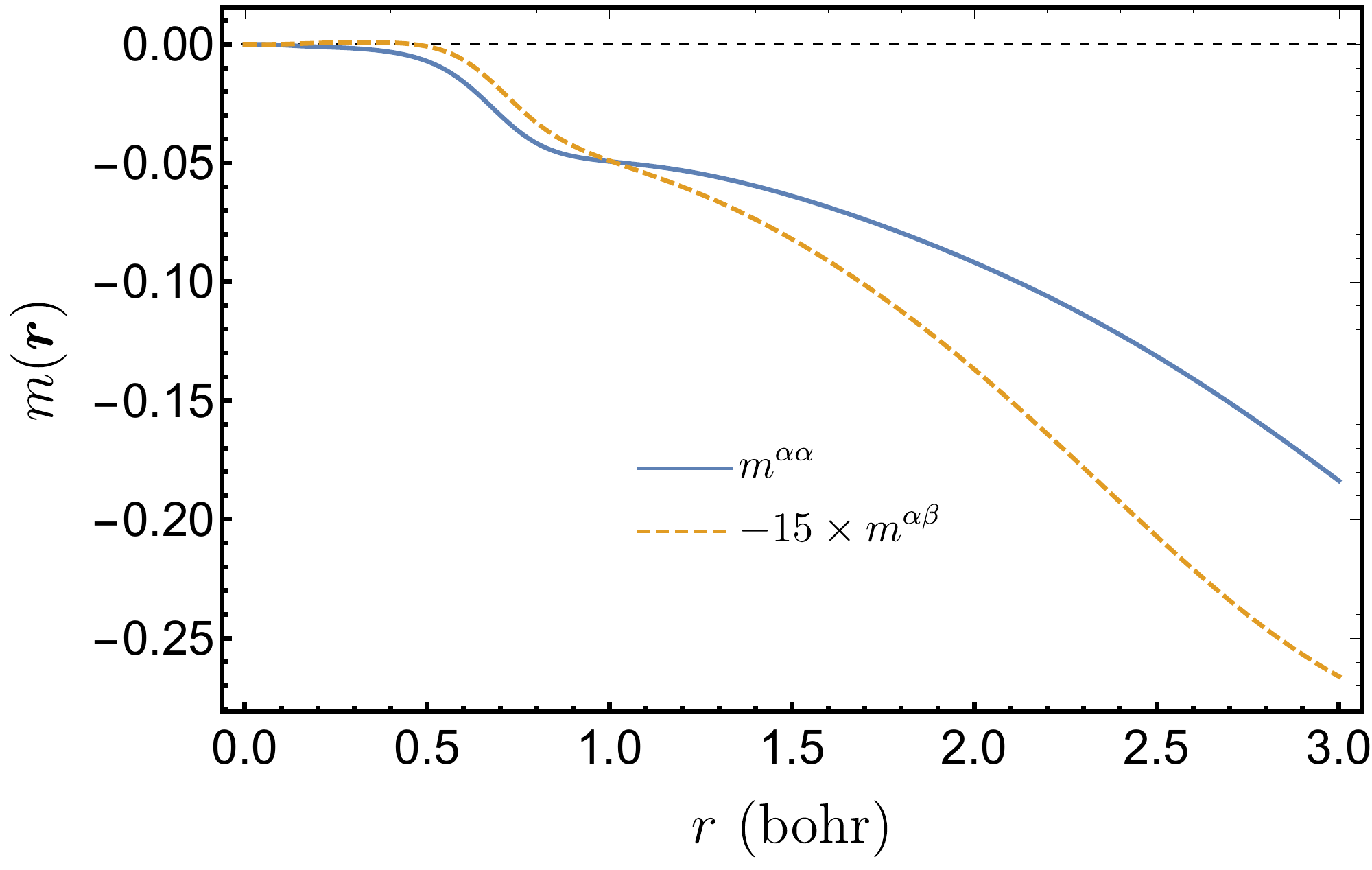}
	\caption{\label{fig:ArrMBD} Comparison between $m^{\alpha\alpha}(\br)$ and scaled $m^{\alpha\beta}(\br)$ of \ce{Ar}.}
	\end{center}
\end{figure}

%-----------------------------------
\label{subsec:comp}
As previously mentioned, the effect of Fermi correlation on relative momenta is orders of magnitude larger than that of Coulomb correlation between parallel-spin electrons.  The effect of Coulomb correlation on opposite-spin electrons is not as small as that on parallel-spin electrons, but there is a difference in magnitude compared to Fermi correlation.  There is also a difference in the relative direction of the correlated momenta; Coulomb correlation of opposite-spin electrons favours equimomentum, whereas Fermi correlation favours antimomentum.  Figures \ref{fig:NerMBD} and \ref{fig:ArrMBD} compare the reduced MB density for \ce{Ne} and \ce{Ar} by scaling the opposite-spin density so that it is similar in magnitude to that of the parallel-spin density, and by reversing the sign. 
It is seen for both \ce{Ne} and \ce{Ar}, that while $m^{\alpha\alpha}(\br)$ and $m^{\alpha\beta}(\br)$ differ in sign, and by an order of magnitude, their shapes are quite similar.  Both $m^{\alpha\alpha}(\br)$ and $m^{\alpha\beta}(\br)$ have minimum amplitude at the nucleus and are relatively flat in the core region.  Both increase in magnitude beyond the core and have a slight shoulder as $r$ increases from the core region to the valence region.  In the valence region, both increase in amplitude towards the extent of appreciable electron density.  At large $r$, the $m(\br)$ begin to differ more. For instance, the $m^{\alpha\beta}(\br)$ of \ce{Ne} approaches a maximum near $r=3$, while $m^{\alpha\alpha}(\br)$ continues to increase in amplitude.  However, the striking similarities of these densities, which are intimately connected to how opposite-spin and parallel-spin electrons avoid each other, cannot be overlooked.

%\FloatBarrier
%---------------------------
\section{Conclusions}
%---------------------------
\label{sec:conc}

A more thorough understanding of the correlated motion of electrons ({\it i.e.~}how electrons avoid each other) is key to unravelling the complexity of the $N$-electron wave function.  Such understanding not only allows for the conception of new, more effective, models of electronic structure, but it also increases our basic knowledge of quantum mechanics.  The equimomentum, antimomentum and momentum-balance, along with their corresponding densities, provide a means for acheiving such understanding. In addition to being a completely new descriptor of the motion of electrons, MB also serves as a correlated motion detector.

The MB density, $\mu(\br)$, povides a measure of the local contribution to momentum-balance in a system of electrons. Due to the dependence of $\mu(\br)$ on the probability an electron is found at $\br$, it resembles the one-electron density.  The reduced MB density, $m(\br)$, removes this dependence and amplifies the more subtle aspects of the local contributions to MB.

The MB, MB density, and reduced MB density can be decomposed into spin contributions. The analysis of the opposite-spin and parallel-spin MB, MB density, and reduced MB density of the atoms \ce{He} to \ce{Ar} has revealed significant differences between the correlated motion of opposite-spin and parallel-spin electrons.  Fermi correlation, or exchange, causes parallel-spin electrons to move with opposite momentum more than equal momentum. For electrons of opposite spin, Coulomb correlation causes them to move with equal momentum more than opposite momentum.  The amount of correlated motion between parallel-spin electrons is strongly dependent on the angular momentum of the electrons present.  Whereas, for opposite-spin electrons, the amount of correlated motion is dependent on simply the number of valence electrons present.  Also, the effect of Fermi correlation on relative electron momenta is an order of magnitude larger than that of Coulomb correlation.

These new tools, particularly the reduced MB density, have also revealed some subtle similarities between the correlated motion of parallel-spin and opposite-spin electrons. For both, correlated motion, given an electron is present, is a minimum near the nucleus and grows as the distance from the nucleus is increased.

The quantities presented, their derivations, and practical algorithms for their calculation are generalized to abitrary systems of multiple nuclei.  Therefore, a natural next step is the investigation of correlated electron motion in molecules.  In addition, because MB and the MB densities reveal the presence of electron correlation, they are useful tools for understanding the role of electron correlation in different chemical phenomena ({\it e.g.}~bond formation/breakage, excited states, dispersion).  They also provide a useful tool for assessing the quality of various models of electronic structure.  These are just a few of the potential applications of these unique quantum mechanical properties.

%-----------------------------------
\section*{Supplementary Material}
%-----------------------------------
See supplementary material for recurrence relations for integrals over Gaussian-type basis functions, and additional equations related to the algorithm for the computation of equimomentum, antimomentum, momentum-balance and their corresponding densities. Also, see supplementary material for plots of $\beta\alpha$ and $\beta\beta$ MB densities, as well as numerical values for total atomic MB. 

%-----------------------------------
\begin{acknowledgements}
%-----------------------------------
JWH thanks the Natural Sciences and Engineering Research Council of Canada (NSERC) for a Discovery Grant, Compute/Calcul Canada for computing resources and the Discovery Institute for Computation and Synthesis for useful consultations.
\end{acknowledgements}

%-----------------------------------
\section*{Data Availability}
%-----------------------------------
The data that support the findings of this study are available from the corresponding author upon request.

%--------------------------
%\appendix
%--------------------------

%--------------------------
%\section{Recurrence Relations}
%--------------------------
%\label{app:RRs}

\bibliography{MBD_atoms}

\pagebreak
\begin{widetext}
\begin{center}
{\textbf{\large Supplementary Material: Measuring correlated electron motion in atoms with the momentum-balance density}}
\end{center}
\end{widetext}

\setcounter{equation}{0}
\setcounter{figure}{0}
\setcounter{table}{0}
\setcounter{page}{1}
\makeatletter
\renewcommand{\theequation}{S\arabic{equation}}
\renewcommand{\thefigure}{S\arabic{figure}}

\newcommand{\ba}{\mathbf{a}}
\newcommand{\bb}{\mathbf{b}}
\newcommand{\bc}{\mathbf{c}}
\newcommand{\bd}{\mathbf{d}}
\newcommand{\bA}{\mathbf{A}}
\newcommand{\bB}{\mathbf{B}}
\newcommand{\bC}{\mathbf{C}}
\newcommand{\bD}{\mathbf{D}}
\newcommand{\bR}{\mathbf{R}}
\newcommand{\bP}{\mathbf{P}}
\newcommand{\bz}{\mathbf{0}}
\newcommand{\lp}{\lambda^+}

%--------------------------
\section*{S.I Recurrence relations for $\lambda^+$}
%--------------------------
The equi- and antimomentum, in terms of the spinless 2-RDM, $\Gamma$, are given by,
\begin{equation}
\label{eq:equiantimom}
\lambda^\pm = \frac{1}{(2\pi)^3} \int \Gamma(\br_1,\br_2,\br_1+\bq,\br_2 \mp \bq)  d\bq d\br_1 d\br_2.
\end{equation}
Expanding the 2-RDM over a general set of basis functions, $\{ \phi_r \}$, leads to the alternative expression for the equi- and antimomentum,
\begin{equation}
\label{eq:equianticont}
\lambda^\pm = \sum_{rstu} \Gamma_{rstu} \left[rstu\right]_{\lambda^\pm},
\end{equation}
where the two-electron integral over four basis functions is given by,
\begin{multline}
	\left[rstu\right]_{\lambda^\pm} = \frac{1}{(2\pi)^3} \int \phi_r(\br_1) \phi_s(\br_1+\bq) \\ 		\times \phi_t(\br_2) \phi_u(\br_2 \mp \bq) d\bq d\br_1 d\br_2.
\end{multline}
In the case of a Gaussian basis set, the fundamental $\lambda^+$ integral\cite{Gill1991,Gill1994} over $s$-type functions with exponents $\alpha$, $\beta$, $\gamma$, $\delta$ and centers $\bA$, $\bB$, $\bC$, $\bD$ is given by,
\begin{equation}
\label{eq:fund}
\left[\bz\bz\bz\bz\right]_{\lambda^+} = \frac{\pi^{3/2} e^{-\sigma R^2}}
{8 \left( \alpha \beta \gamma + \beta \gamma \delta + \alpha \beta \delta + \alpha \gamma \delta \right)^{3/2}},
\end{equation}
where
\begin{equation}
	\sigma =  \frac{\alpha \beta \gamma \delta}
{\alpha \beta \gamma + \beta \gamma \delta + \alpha \beta \delta + \alpha \gamma \delta}
\end{equation}
and
\begin{equation}
	\bR = \bA + \bC - \bB - \bD.
\end{equation}
Integrals of higher angular momentum can be constructed using a five-term recurrence relation (RR).\cite{RR11} The RRs for augmenting the angular momentum of the Gaussians on each centre, $\bA$, $\bB$, $\bC$ and $\bD$, in the $i$-direction (where $i =x,y$ or $z$) are given by
\begin{subequations}
\begin{align}
% equi a+1
	[(\ba + \bm{1}_i)\bb \bc \bd]_{\lp} = & 
	-\frac{\sigma}{\alpha}R_i [\ba \bb \bc \bd]_{\lp} \nonumber \\
	&+ a_i\frac{\alpha - \sigma}{2\alpha^2} [(\ba - \bm{1}_i)\bb \bc \bd]_{\lp} \nonumber \\
	&+ b_i\frac{\sigma}{2\alpha\beta} [\ba (\bb - \bm{1}_i) \bc \bd]_{\lp}  \nonumber \\
	&- c_i\frac{\sigma}{2\alpha\gamma} [\ba \bb (\bc - \bm{1}_i) \bd]_{\lp} \nonumber \\
	&+ d_i\frac{\sigma}{2\alpha\delta} [\ba \bb \bc (\bd - \bm{1}_i)]_{\lp},
\end{align}
\begin{align}
% equi b+1
	[\ba (\bb + \bm{1}_i) \bc \bd]_{\lp} = &
	\frac{\sigma}{\beta}R_i [\ba \bb \bc \bd]_{\lp} \nonumber \\
	&+ a_i\frac{\sigma}{2\alpha\beta} [(\ba - \bm{1}_i)\bb \bc \bd]_{\lp} \nonumber \\
	&+ b_i\frac{\beta - \sigma}{2\beta^2} [\ba (\bb - \bm{1}_i) \bc \bd]_{\lp} \nonumber \\
	&+ c_i\frac{\sigma}{2\beta\gamma} [\ba \bb (\bc - \bm{1}_i) \bd]_{\lp} \nonumber \\
	&- d_i\frac{\sigma}{2\beta\delta} [\ba \bb \bc (\bd - \bm{1}_i)]_{\lp},
\end{align}
\begin{align}
% equi c+1
	[\ba \bb (\bc + \bm{1}_i) \bd]_{\lp} = &
	-\frac{\sigma}{\gamma}R_i [\ba \bb \bc \bd]_{\lp} \nonumber \\
	&- a_i\frac{\sigma}{2\alpha\gamma} [(\ba - \bm{1}_i)\bb \bc \bd]_{\lp} \nonumber \\
	&+ b_i\frac{\sigma}{2\beta\gamma} [\ba (\bb - \bm{1}_i) \bc \bd]_{\lp} \nonumber \\
	&+ c_i\frac{\gamma - \sigma}{2\gamma^2} [\ba \bb (\bc - \bm{1}_i) \bd]_{\lp} \nonumber \\
	&+ d_i\frac{\sigma}{2\gamma\delta} [\ba \bb \bc (\bd - \bm{1}_i)]_{\lp},
\end{align}
\begin{align}
% equi d+1
	[\ba \bb \bc (\bd + \bm{1}_i)]_{\lp} = &
	\frac{\sigma}{\delta}R_i [\ba \bb \bc \bd]_{\lp} \nonumber \\
	&+ a_i\frac{\sigma}{2\alpha\delta} [(\ba - \bm{1}_i)\bb \bc \bd]_{\lp} \nonumber \\
	&- b_i\frac{\sigma}{2\beta\delta} [\ba (\bb - \bm{1}_i) \bc \bd]_{\lp} \nonumber \\
	&+ c_i\frac{\sigma}{2\gamma\delta} [\ba \bb (\bc - \bm{1}_i) \bd]_{\lp} \nonumber \\
	&+ d_i\frac{\delta - \sigma}{2\delta^2} [\ba \bb \bc (\bd - \bm{1}_i)]_{\lp},
\end{align}
\end{subequations}
where $\ba=(a_x,a_y,a_z)$ is a vector of angular momentum quantum numbers, and $\bm{1}_i$ is a unit vector with value 1 in the $i$-direction [{\it e.g.}~$\bm{1}_x =(1,0,0)$].

The momentum-balance (MB) is the difference between the equi- and antimomentum,
\begin{equation}
	\label{eq:MB}
	\mu = \lambda^+ - \lambda^-.
\end{equation}
The effort required to calculate MB is halved by making use of the following identity, 
\begin{equation}
	\label{eq:ident}
	[rstu]_{\lambda^-} = [rsut]_{\lp}.
\end{equation}
Insertion of the 2-RDM expansion (Equation \ref{eq:equianticont}) into Equation \ref{eq:MB} and use of the identity above (Equation \ref{eq:ident}) leads to a practical equation for the calculation of MB,
\begin{equation}
	\mu = \sum_{rstu} \Gamma_{rstu} \left([rstu]_{\lp} - [rsut]_{\lp} \right).
\end{equation}

The equi- and antimomentum densities are given by the following equation,
\begin{equation}
\label{eq:equiantiden}
\lambda^\pm(\br) = \frac{1}{(2\pi)^3} \int \Gamma(\br,\br_2,\br+\bq,\br_2 \mp \bq)  d\bq d\br_2.
\end{equation}
Similar to equi- and antimomentum, their corresponding densities can also be expanded over a basis set,
\begin{equation}
\label{eq:equianticont}
\lambda^\pm(\br) = \sum_{rstu} \Gamma_{rstu} \phi_r(\br)\left[stu\right]_{\lambda^\pm},
\end{equation}
where the three-function integral is given by,
\begin{multline}
	\left[stu\right]_{\lambda^\pm} = \frac{1}{(2\pi)^3} \int \phi_s(\br+\bq) \\ 		\times \phi_t(\br_2) \phi_u(\br_2 \mp \bq) d\bq d\br_2.
\end{multline}
If the basis set is comprised of Gaussian-type functions, then the fundamental $\lambda^+(\br)$ integral is given by,
\begin{equation}
\label{eq:fund}
\left[\bz\bz\bz\right]_{\lambda^+} = \frac{e^{-\tau P^2}}
{8 \left( \beta \gamma + \beta \delta + \gamma \delta \right)^{3/2}},
\end{equation}
where
\begin{equation}
	\tau =  \frac{\beta \gamma \delta}{\beta \gamma + \beta \delta + \gamma \delta},
\end{equation}
and
\begin{equation}
	\bP = \br - \bB + \bC - \bD.
\end{equation}
The integrals over functions of higher angular momentum can be calculated using the following 4-term recurrence relations,
\begin{subequations}
\begin{align}
% equi b+1
	[(\bb + \bm{1}_i) \bc \bd]_{\lp} = &
	\frac{\tau}{\beta}P_i [\bb \bc \bd]_{\lp} \nonumber \\
	&+ b_i\frac{\beta - \tau}{2\beta^2} [(\bb - \bm{1}_i) \bc \bd]_{\lp} \nonumber \\
	&+ c_i\frac{\tau}{2\beta\gamma} [\bb (\bc - \bm{1}_i) \bd]_{\lp} \nonumber \\
	&- d_i\frac{\tau}{2\beta\delta} [\bb \bc (\bd - \bm{1}_i)]_{\lp},
\end{align}
\begin{align}
% equi c+1
	[\bb (\bc + \bm{1}_i) \bd]_{\lp} = &
	-\frac{\tau}{\gamma}P_i [\bb \bc \bd]_{\lp} \nonumber \\
	&+ b_i\frac{\tau}{2\beta\gamma} [(\bb - \bm{1}_i) \bc \bd]_{\lp} \nonumber \\
	&+ c_i\frac{\gamma - \tau}{2\gamma^2} [\bb (\bc - \bm{1}_i) \bd]_{\lp} \nonumber \\
	&+ d_i\frac{\tau}{2\gamma\delta} [\bb \bc (\bd - \bm{1}_i)]_{\lp},
\end{align}
\begin{align}
% equi d+1
	[\bb \bc (\bd + \bm{1}_i)]_{\lp} = &
	\frac{\tau}{\delta}P_i [\bb \bc \bd]_{\lp} \nonumber \\
	&- b_i\frac{\tau}{2\beta\delta} [(\bb - \bm{1}_i) \bc \bd]_{\lp} \nonumber \\
	&+ c_i\frac{\tau}{2\gamma\delta} [\bb (\bc - \bm{1}_i) \bd]_{\lp} \nonumber \\
	&+ d_i\frac{\delta - \tau}{2\delta^2} [\bb \bc (\bd - \bm{1}_i)]_{\lp}.
\end{align}
\end{subequations}
The MB density is calculated as the difference between the equi- and antimomentum densities,
\begin{equation}
	\mu(\br) = \lambda^+(\br) - \lambda^-(\br).
\end{equation}
To efficiently calculate $\mu(\br)$, the following identity is used,
\begin{equation}
	[stu]_{\lambda^-} = [sut]_{\lp}.
\end{equation}
If calculating $\mu(\br)$ for numerous values of $\br$, it is desirable to reduce the number of integral evaluations, which can be done with the following expression,
\begin{equation}
	\mu(\br) = \sum_{rstu} \left( \Gamma_{rstu} - \Gamma_{rsut} \right) \phi_i(\br) [stu]_{\lp}.
\end{equation}

%--------------------------
\section*{S.II Atomic momentum-balance densities}
%--------------------------
\FloatBarrier

\begin{figure}
	\begin{center}
	\includegraphics[width=0.48\textwidth]{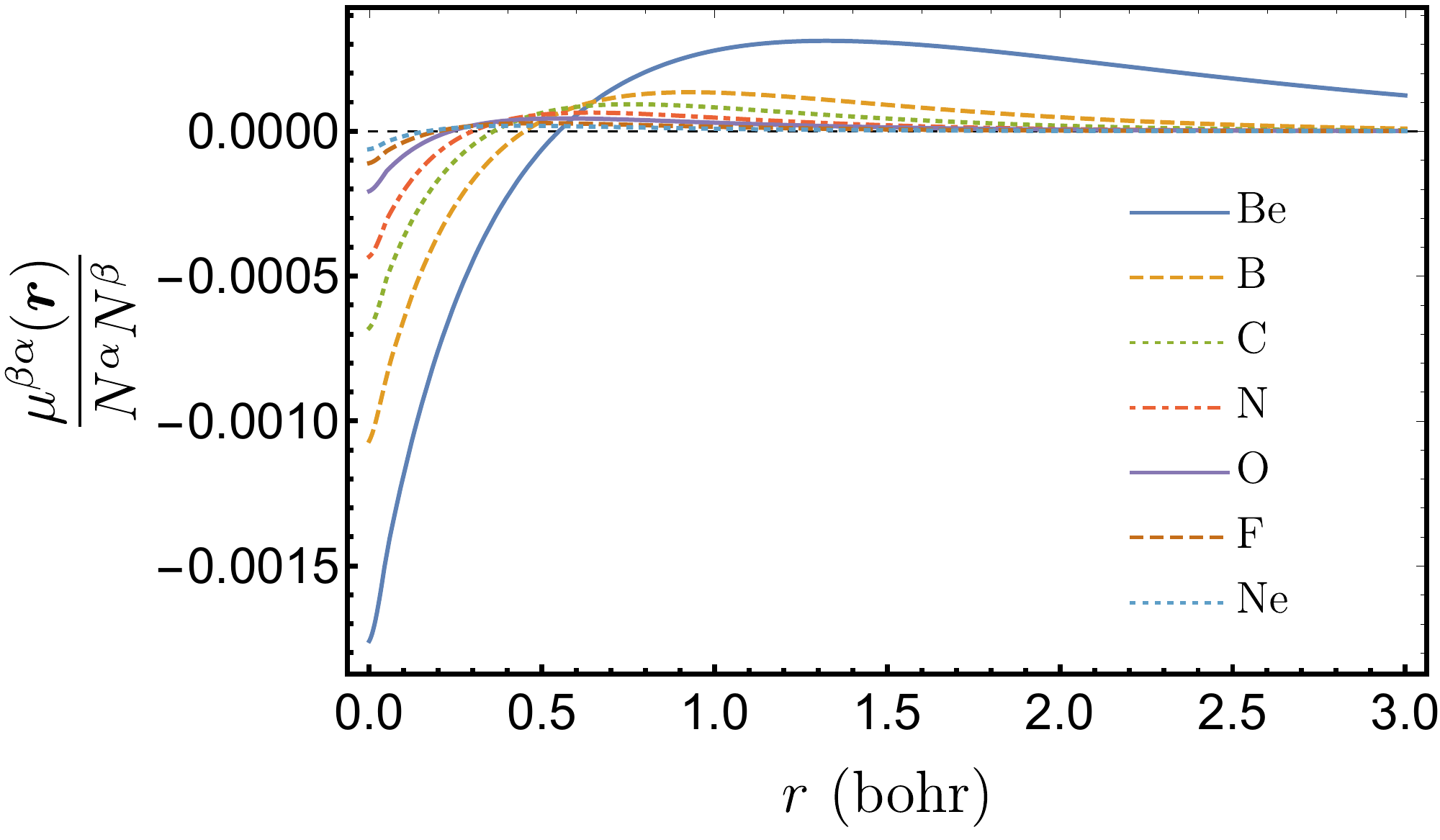}
	\caption{\label{fig:BetoNeMBDba} Renormalized $\mu^{\beta\alpha}(\br)$ of the atoms \ce{Be} to \ce{Ne}.}
	\end{center}
\end{figure}

\begin{figure}
	\begin{center}
	\includegraphics[width=0.48\textwidth]{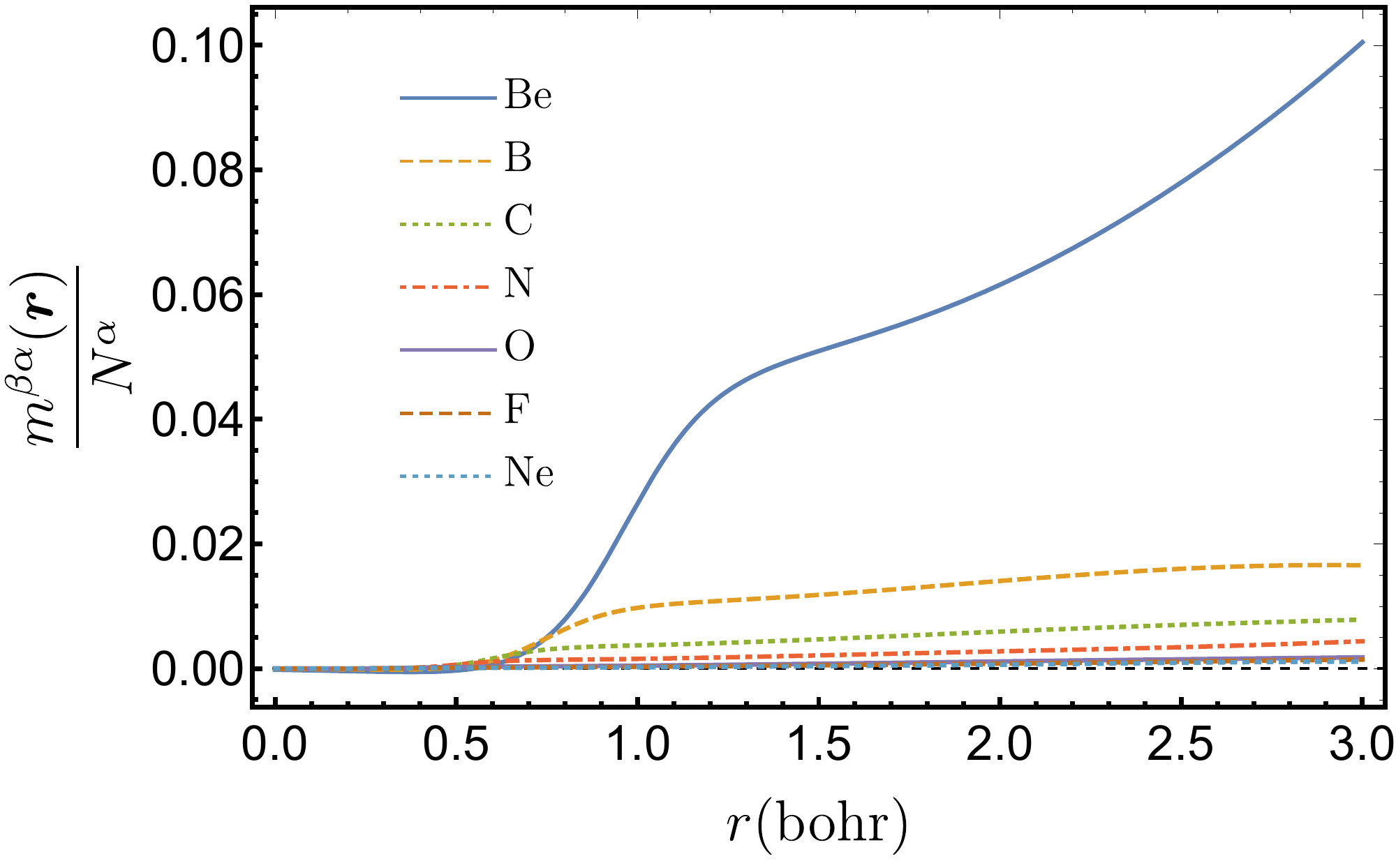} \includegraphics[width=0.48\textwidth]{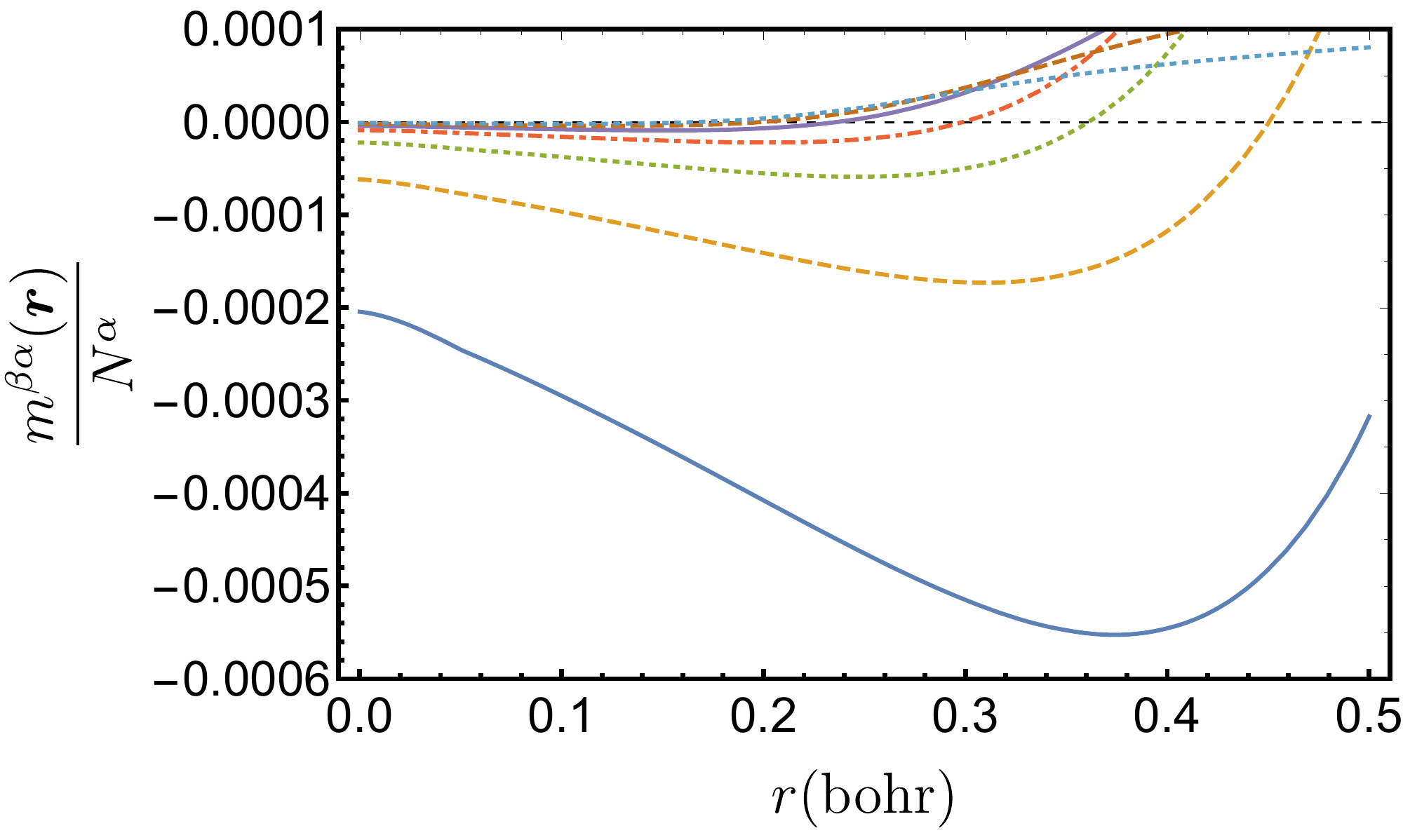}
	\caption{\label{fig:BetoNerMBDba} Renormalized $m^{\beta\alpha}(\br)$ of the atoms \ce{Be} to \ce{Ne}.}
	\end{center}
\end{figure}

\begin{figure}
	\begin{center}
	\includegraphics[width=0.48\textwidth]{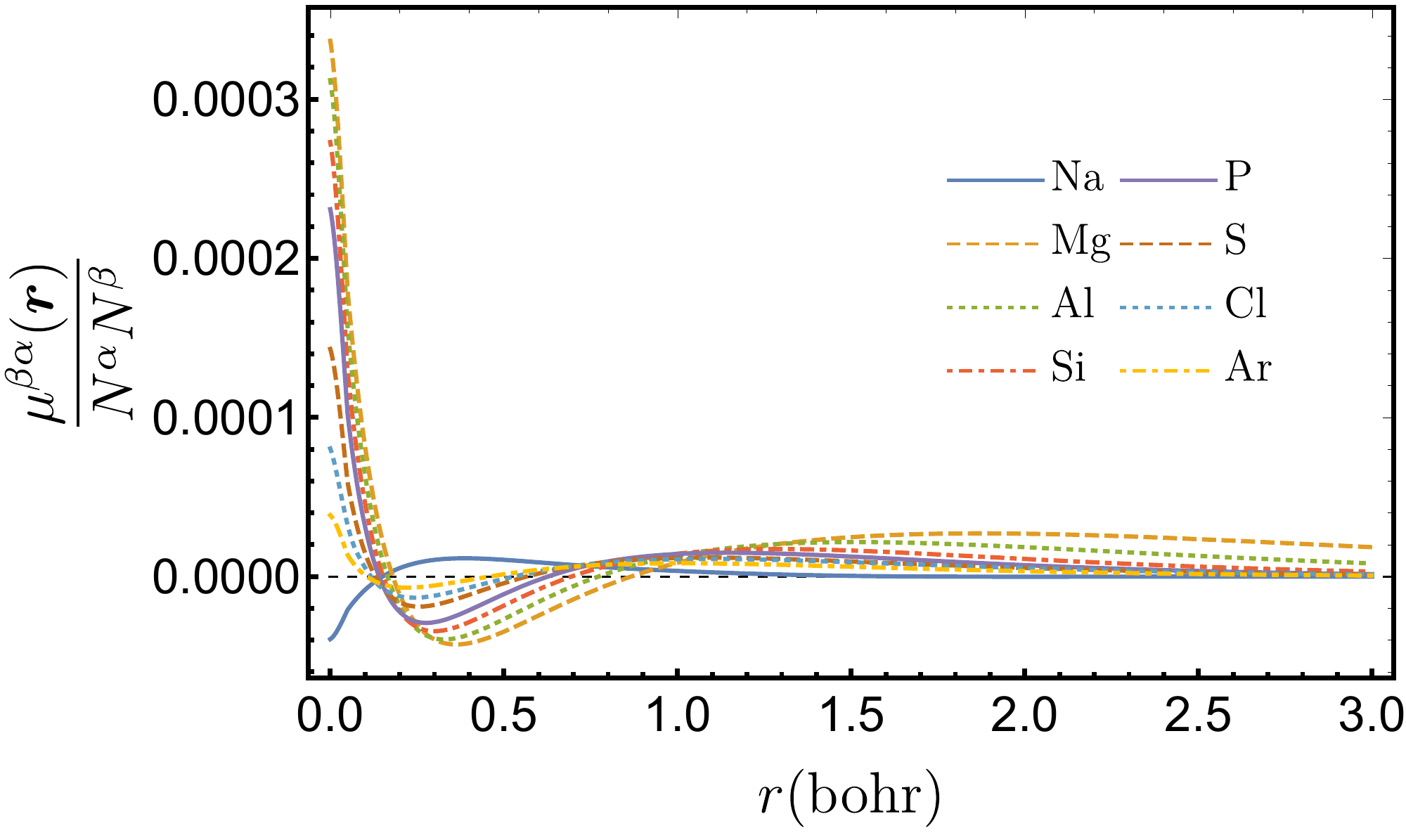}
	\caption{\label{fig:NatoArMBDba} Renormalized $\mu^{\beta\alpha}(\br)$ of the atoms \ce{Na} to \ce{Ar}.}
	\end{center}
\end{figure}

\begin{figure}
	\begin{center}
	\includegraphics[width=0.48\textwidth]{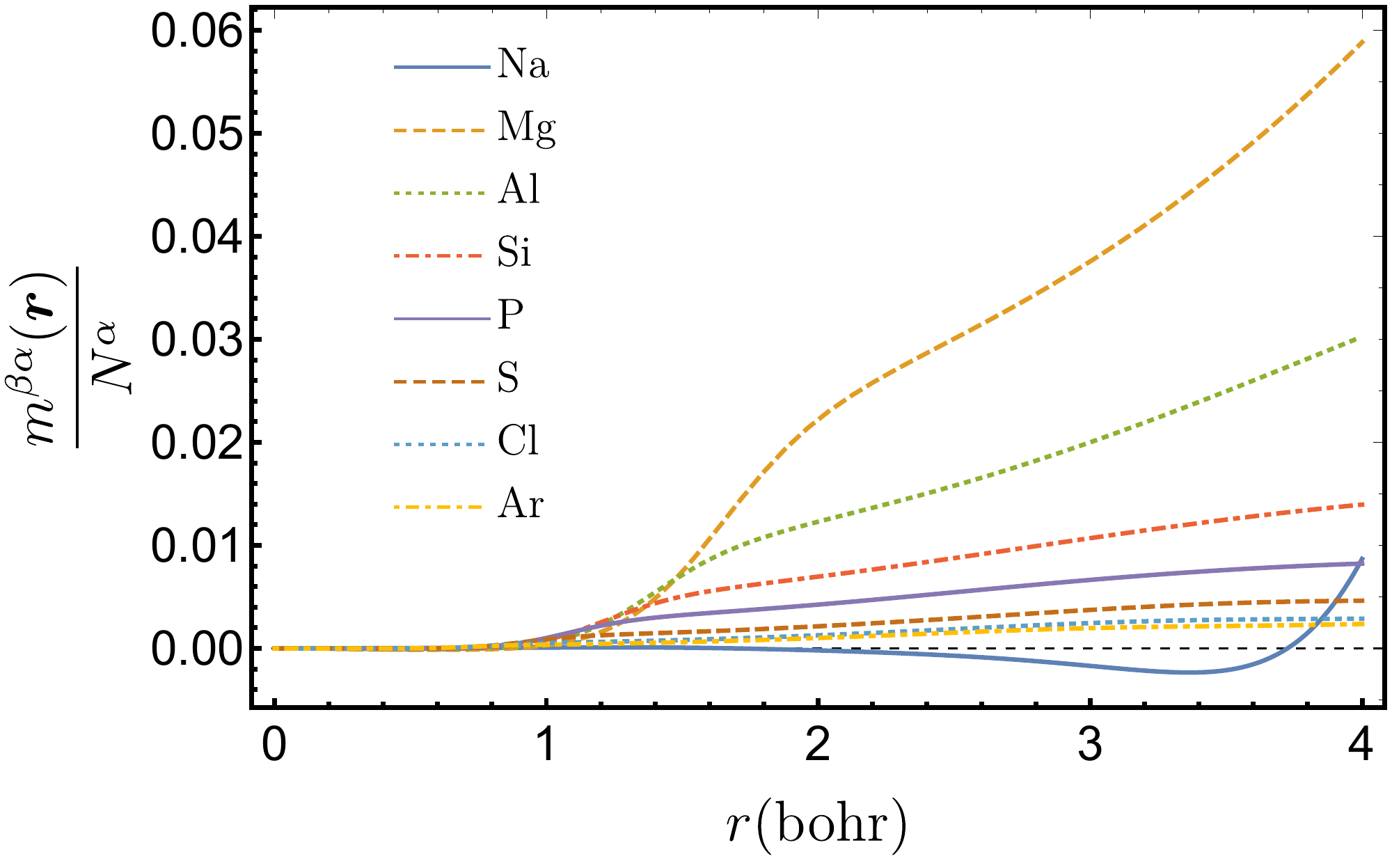} \includegraphics[width=0.48\textwidth]{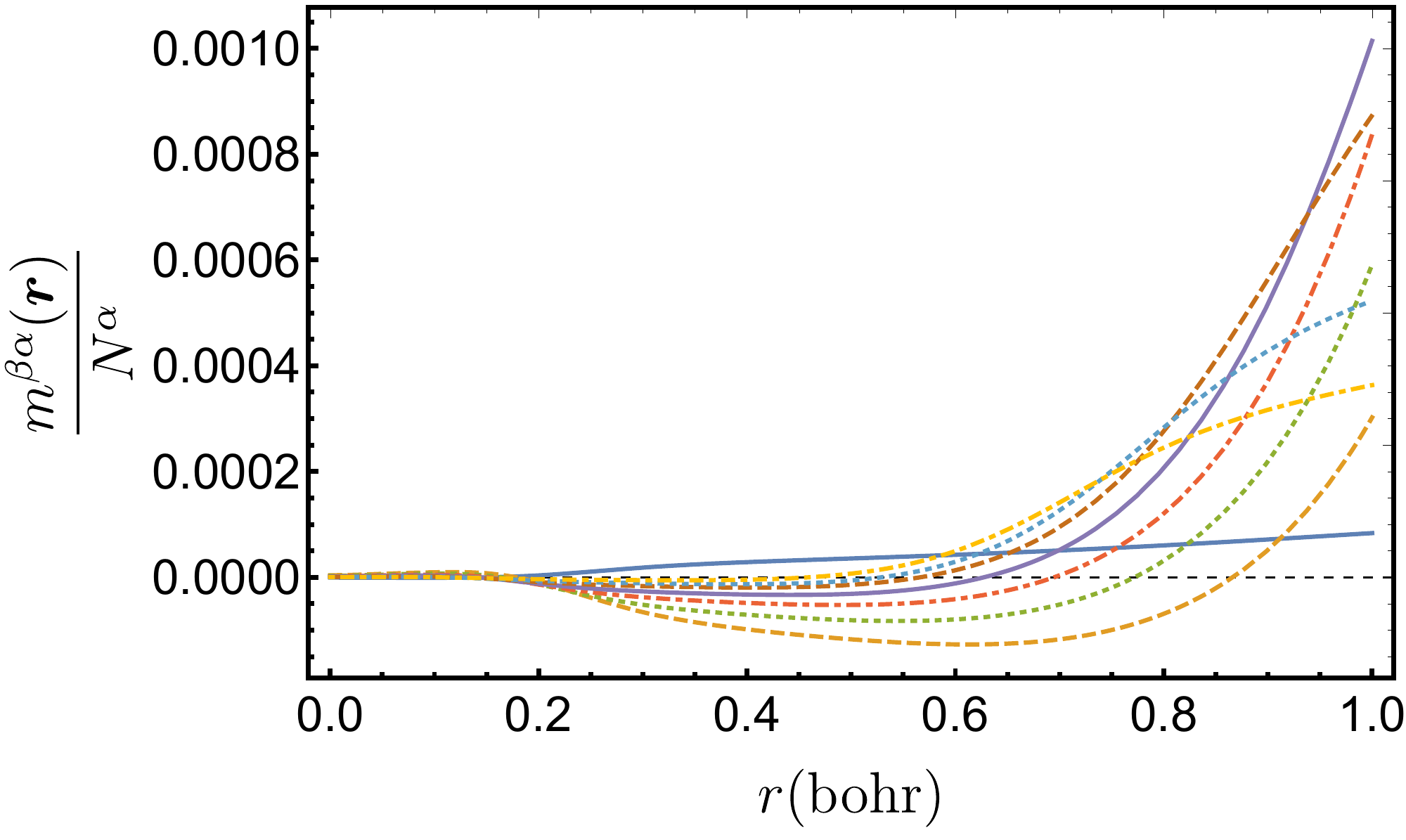} \includegraphics[width=0.48\textwidth]{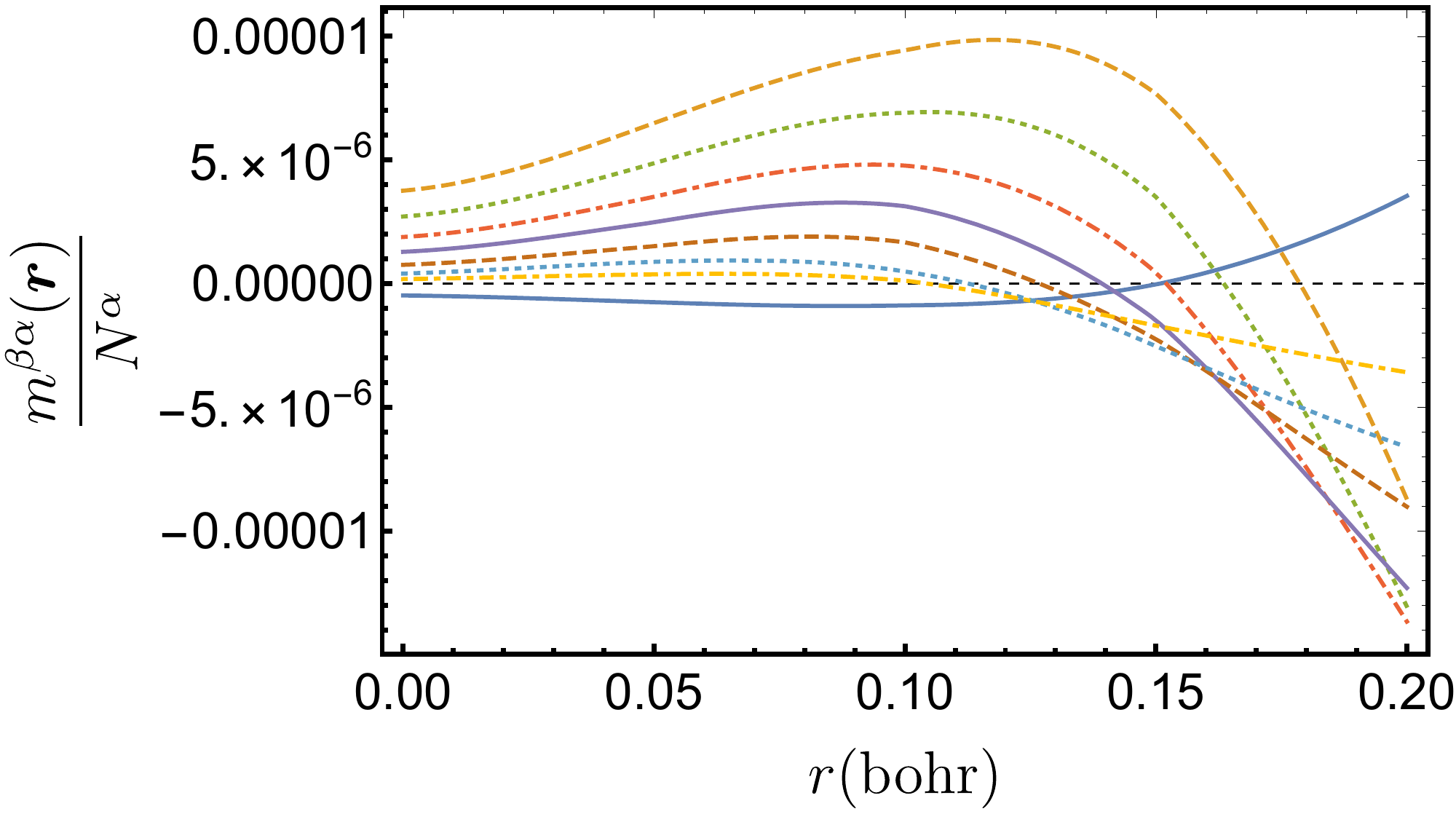}
	\caption{\label{fig:NatoArrMBDba} Renormalized $m^{\beta\alpha}(\br)$ of the atoms \ce{Na} to \ce{Ar}.}
	\end{center}
\end{figure}

\begin{figure}
	\begin{center}
	\includegraphics[width=0.48\textwidth]{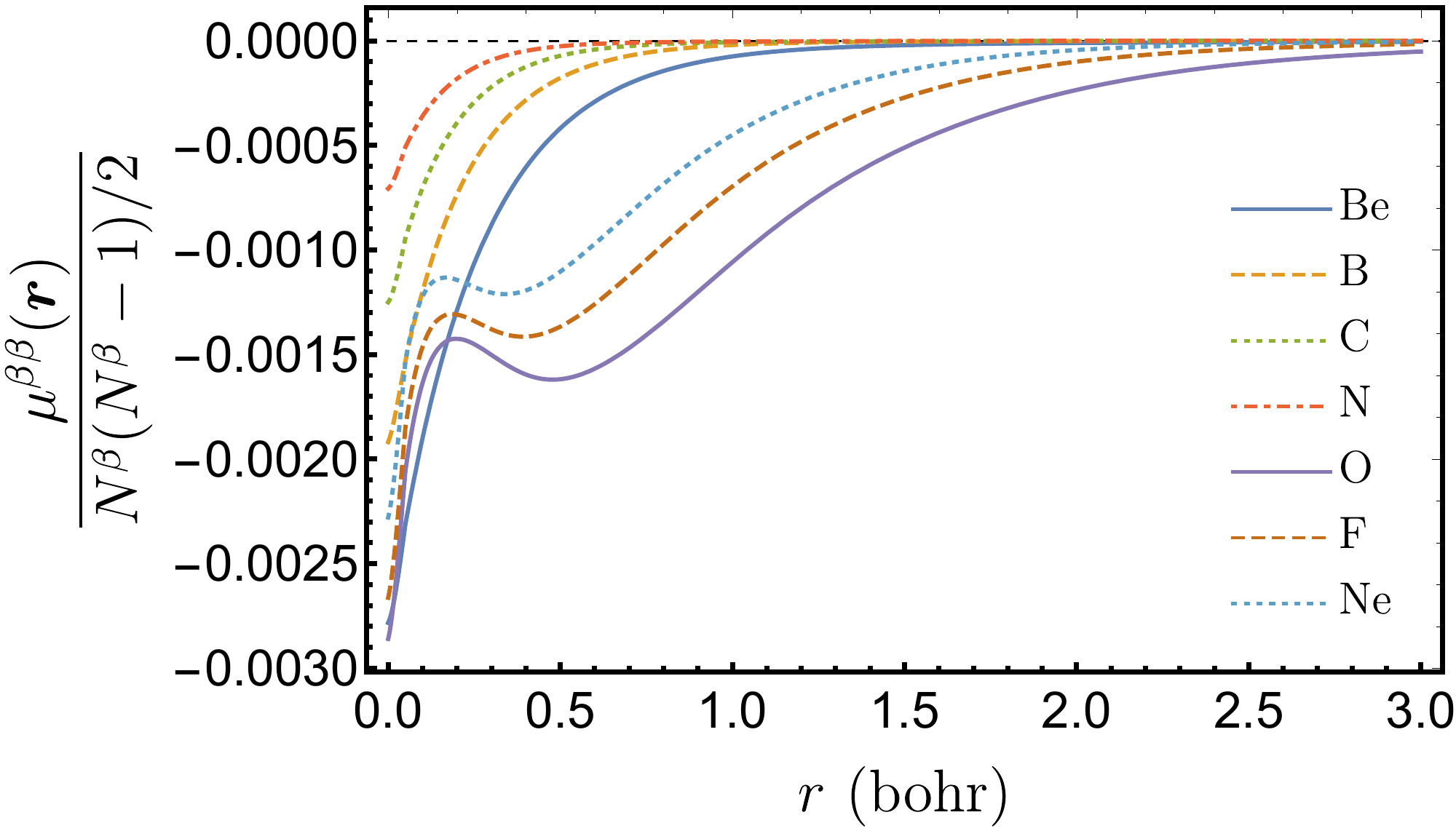}
	\caption{\label{fig:BetoNeMBDbb} Renormalized $\mu^{\beta\beta}(\br)$ of the atoms \ce{Be} to \ce{Ne}.}
	\end{center}
\end{figure}

\begin{figure}
	\begin{center}
	\includegraphics[width=0.48\textwidth]{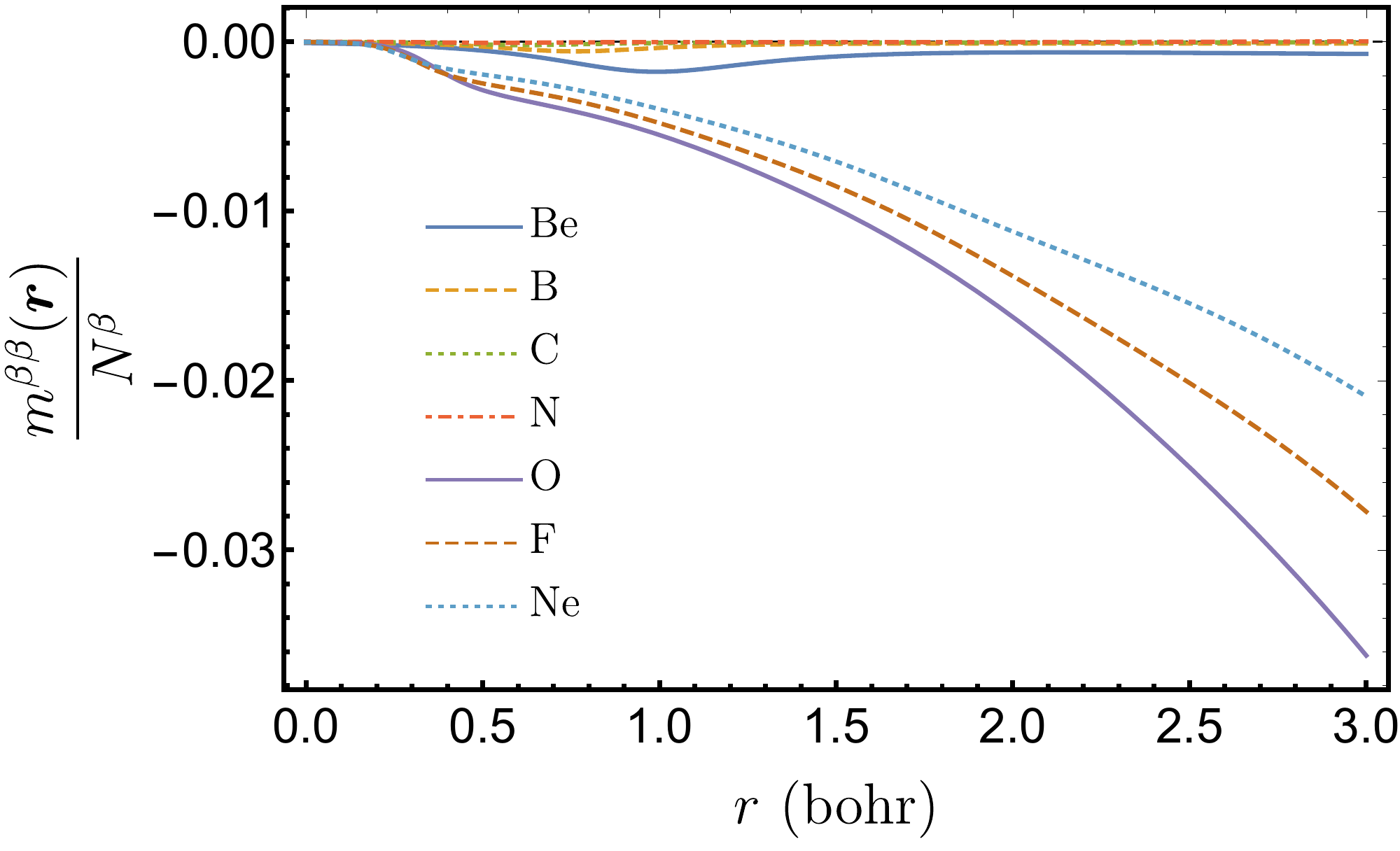}
	\includegraphics[width=0.48\textwidth]{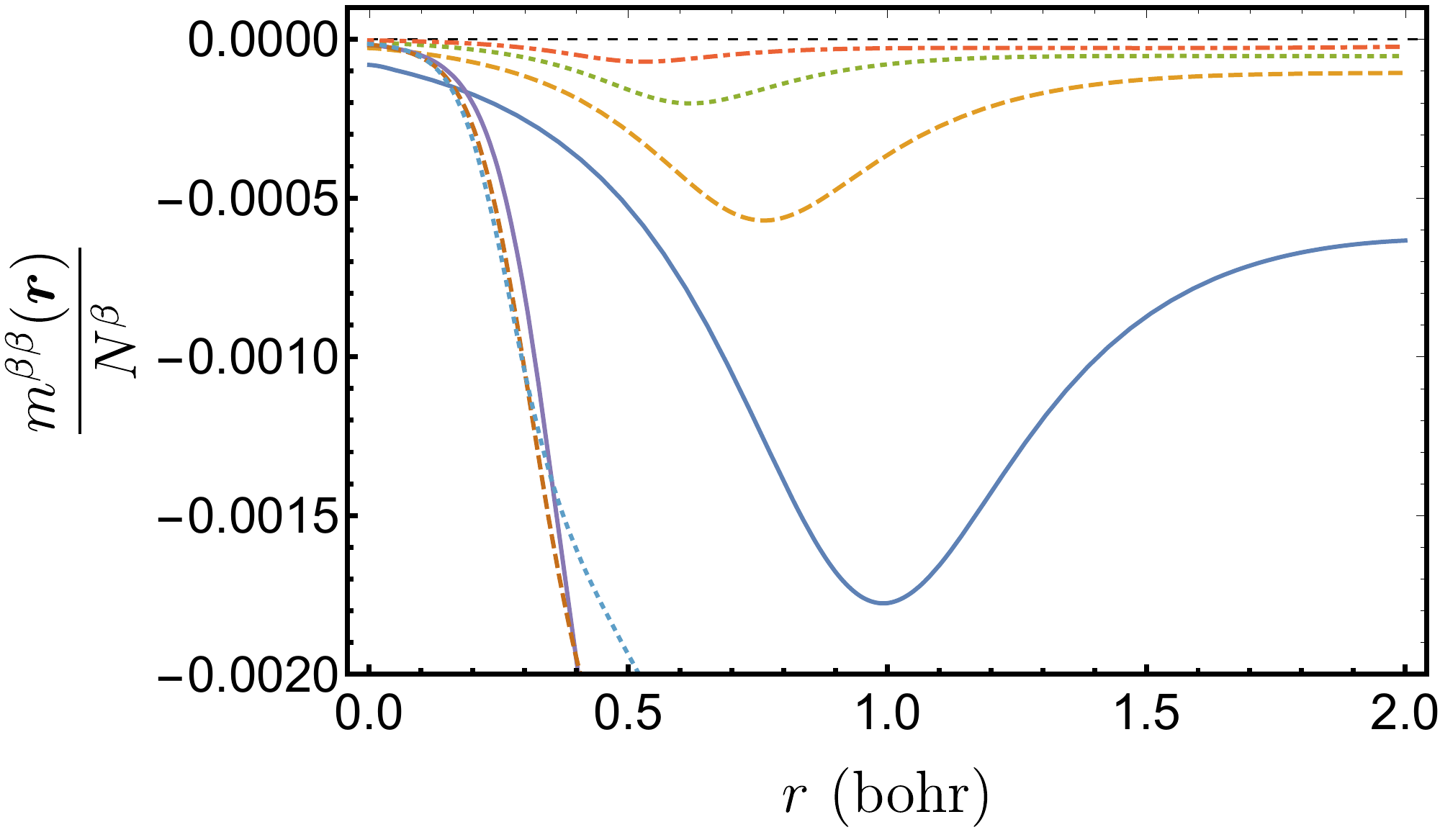}
	\caption{\label{fig:BetoNerMBDbb} Renormalized $m^{\beta\beta}(\br)$ of the atoms \ce{Be} to \ce{Ne}.}
	\end{center}
\end{figure}

\begin{figure}
	\begin{center}
	\includegraphics[width=0.48\textwidth]{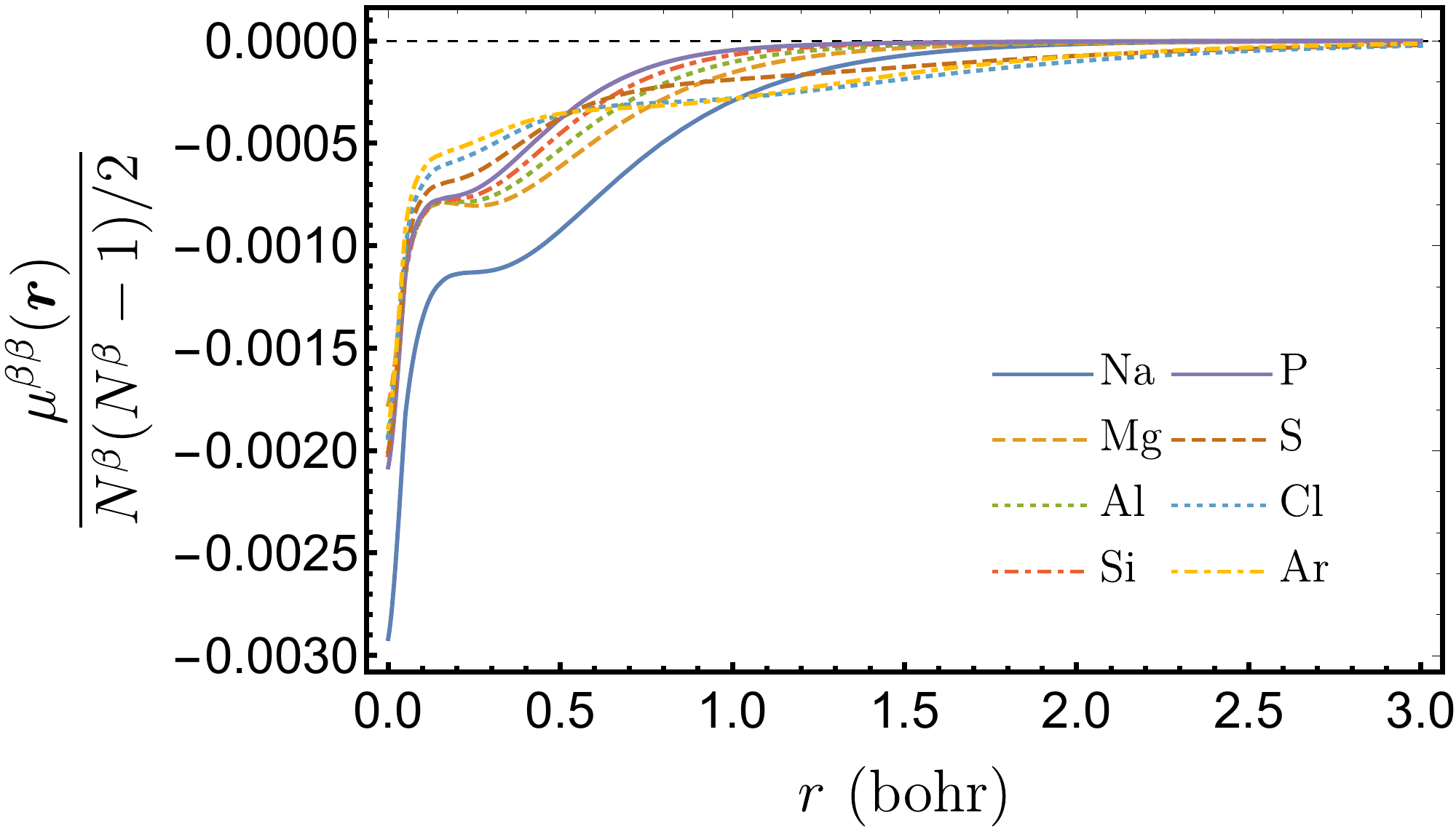}
	\caption{\label{fig:NatoArMBDbb} Renormalized $\mu^{\beta\beta}(\br)$ of the atoms \ce{Na} to \ce{Ar}.}
	\end{center}
\end{figure}

\begin{figure}
	\begin{center}
	\includegraphics[width=0.48\textwidth]{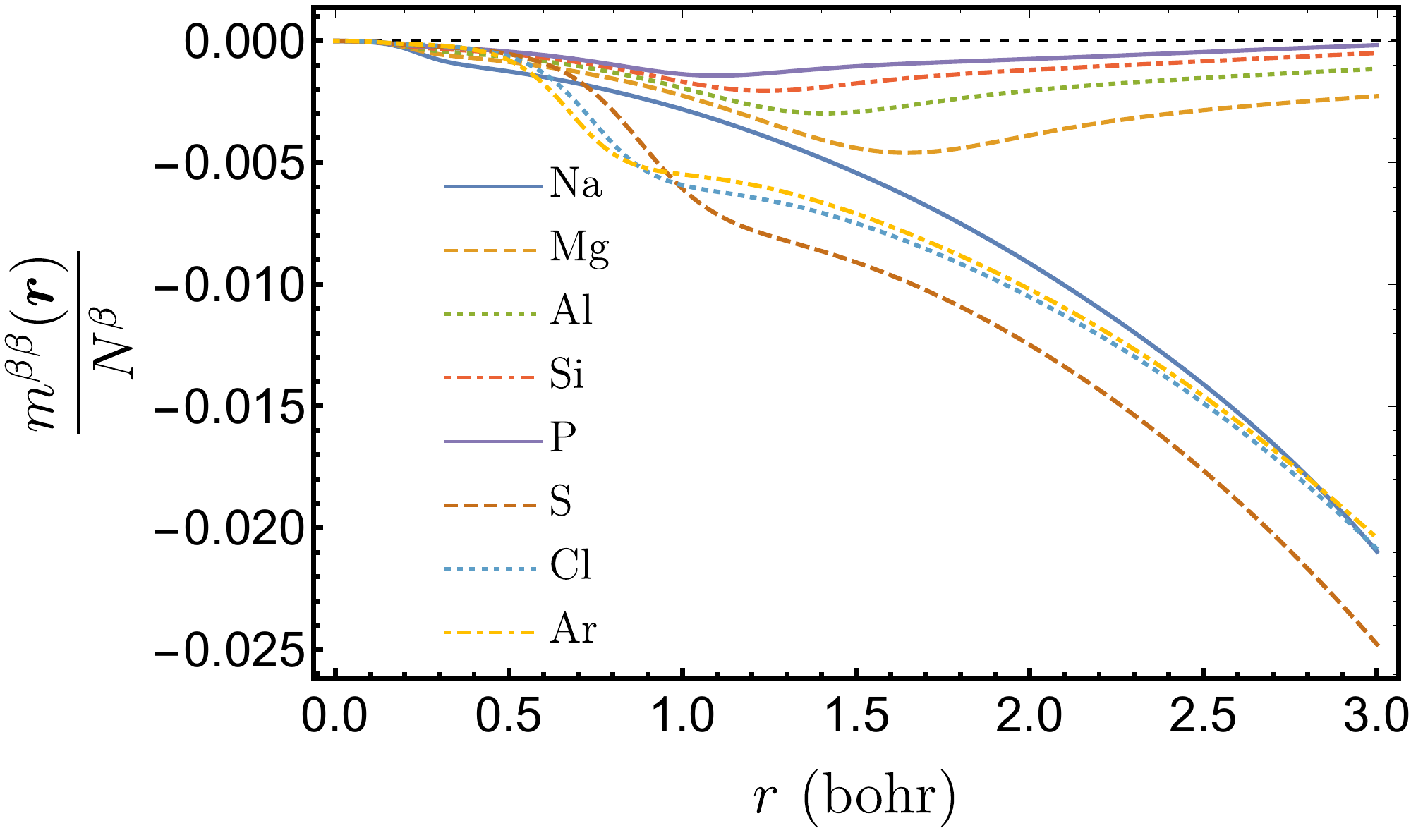} \includegraphics[width=0.48\textwidth]{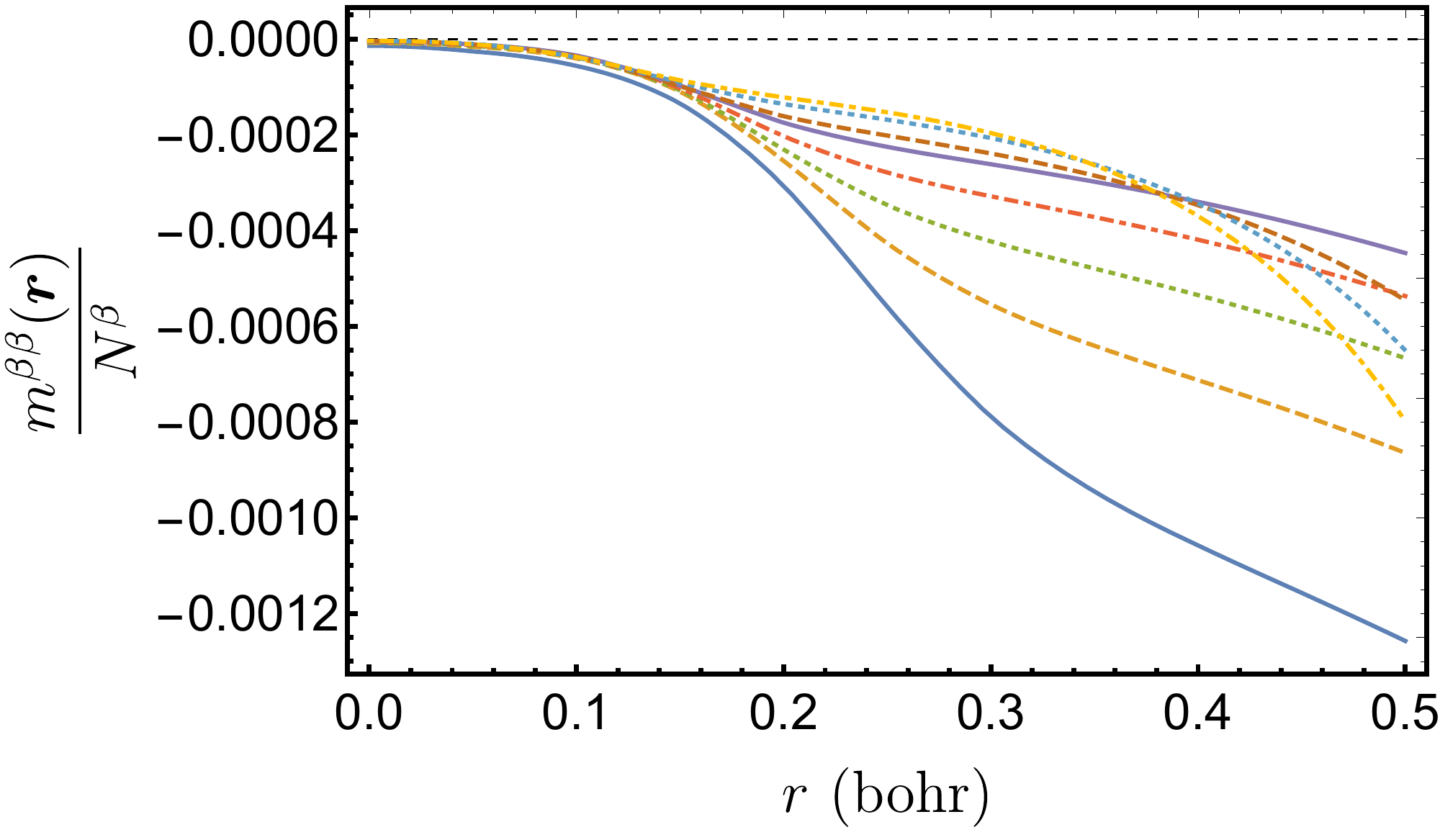}
	\caption{\label{fig:NatoArrMBDbb} Renormalized $m^{\beta\beta}(\br)$ of the atoms \ce{Na} to \ce{Ar}.}
	\end{center}
\end{figure}

The $\mu^{\beta\alpha}(\br)$, $m^{\beta\alpha}(\br)$ and $\mu^{\beta\beta}(\br)$, $m^{\beta\beta}(\br)$ of the atoms of the first three rows of the periodic table are presented in Figures \ref{fig:BetoNeMBDba} to \ref{fig:NatoArrMBDbb}.

\FloatBarrier
%--------------------------
\section*{S.III Atomic momentum-balance}
%--------------------------
\FloatBarrier

The total opposite-spin and parallel-spin momentum-balance for the atoms \ce{He} to \ce{Ar} are reported in Table \ref{tab:MB}.

%\begin{turnpage}
\begin{table}
	\caption{\label{tab:MB} Spin-resolved and total momentum-balance of the atoms helium to argon}
	\begin{ruledtabular}
	\begin{tabular}{lrrrr}
	Atom	&	$\mu^{\alpha\beta}$	&	$\mu^{\alpha\alpha}$	&	$\mu^{\beta\beta}$	&	$\mu$	\\
	\hline
	\ce{He}		& 0.00373	& 0		&  0		&	0.00373		\\
	\ce{Li}		& 0.00107	& -0.00001	&  0		&	0.00212		\\
	\ce{Be}		& 0.19320	& -0.00240	& -0.00240	&	0.38160		\\
	\ce{B}		& 0.06497	& -0.25768	& -0.00077	&	-0.12851	\\
	\ce{C}		& 0.02555	& -0.27478	& -0.00028	&	-0.22396	\\
	\ce{N}		& 0.01029	& -0.24801	& -0.00010	&	-0.22752	\\	
	\ce{O}		& 0.00770	& -0.16412	& -0.06507	&	-0.21380	\\
	\ce{F}		& 0.00564	& -0.11649	& -0.08387	&	-0.18908	\\
	\ce{Ne}		& 0.00419	& -0.08663	& -0.08663	&	-0.16487	\\
	\ce{Na}		& 0.00095	& -0.07077	& -0.04902	&	-0.11788	\\
	\ce{Mg}		& 0.29053	& -0.04855	& -0.04855	&	 0.48398	\\
	\ce{Al}		& 0.13933	& -0.76686	& -0.03289	&	-0.52109	\\
	\ce{Si}		& 0.07348	& -0.89145	& -0.02315	&	-0.76764	\\
	\ce{P}		& 0.03908	& -0.87708	& -0.01680	&	-0.81572	\\
	\ce{S}		& 0.03920	& -0.60966	& -0.21962	&	-0.75088	\\
	\ce{Cl}		& 0.03394	& -0.44856	& -0.30727	&	-0.68796	\\
	\ce{Ar}		& 0.02828	& -0.34316	& -0.34316	&	-0.62976	\\
	\end{tabular}
	\end{ruledtabular}
\end{table}
%\end{turnpage}

\FloatBarrier

%\begin{thebibliography}
%\cite{Gill1991,Gill1994,RR11}
%\end{thebibliography}

\end{document}